\pdfoutput=1
\documentclass[12pt,a4paper]{article}
\usepackage{ifthen} % for conditional statements
\newboolean{pdflatex}
\setboolean{pdflatex}{true} % False for eps figures 

\newboolean{articletitles}
\setboolean{articletitles}{true} % False removes titles in references

\newboolean{uprightparticles}
\setboolean{uprightparticles}{true} %True for upright particle symbols

\newboolean{inbibliography}
\setboolean{inbibliography}{false} %True once you enter the bibliography

\usepackage[normalem]{ulem}

\usepackage{microtype}
\usepackage{lineno}  % for line numbering during review
\usepackage{xspace} % To avoid problems with missing or double spaces after
\usepackage{caption} %these three command get the figure and table captions automatically small

\usepackage{graphicx}  % to include figures (can also use other packages)
\usepackage{color}
\usepackage{colortbl}
\graphicspath{{./figs/}} % Make Latex search fig subdir for figures

\usepackage{amsmath} % Adds a large collection of math symbols
\usepackage{amssymb}
\usepackage{amsfonts}
\usepackage{upgreek} % Adds in support for greek letters in roman typeset

\newcommand*\patchAmsMathEnvironmentForLineno[1]{%
\expandafter\let\csname old#1\expandafter\endcsname\csname #1\endcsname
\expandafter\let\csname oldend#1\expandafter\endcsname\csname
end#1\endcsname
 \renewenvironment{#1}%
   {\linenomath\csname old#1\endcsname}%
   {\csname oldend#1\endcsname\endlinenomath}%
}
\newcommand*\patchBothAmsMathEnvironmentsForLineno[1]{%
  \patchAmsMathEnvironmentForLineno{#1}%
  \patchAmsMathEnvironmentForLineno{#1*}%
}
\AtBeginDocument{%
\patchBothAmsMathEnvironmentsForLineno{equation}%
\patchBothAmsMathEnvironmentsForLineno{align}%
\patchBothAmsMathEnvironmentsForLineno{flalign}%
\patchBothAmsMathEnvironmentsForLineno{alignat}%
\patchBothAmsMathEnvironmentsForLineno{gather}%
\patchBothAmsMathEnvironmentsForLineno{multline}%
\patchBothAmsMathEnvironmentsForLineno{eqnarray}%
}

\usepackage{hyperref}    % Hyperlinks in references
\usepackage[all]{hypcap} % Internal hyperlinks to floats.

\usepackage{xspace} 
\usepackage{upgreek}

\def\lhcb {\mbox{LHCb}\xspace}
\def\atlas  {\mbox{ATLAS}\xspace}

\def\babar  {\mbox{BaBar}\xspace}

\def\cdf    {\mbox{CDF}\xspace}

\def\MagUp {\mbox{\em Mag\kern -0.05em Up}\xspace}

\ifthenelse{\boolean{uprightparticles}}%
{

 \def\Pmu         {\ensuremath{\upmu}\xspace}

 \def\Ppi         {\ensuremath{\uppi}\xspace}

 \def\Ppsi        {\ensuremath{\uppsi}\xspace}

 \def\PDelta      {\ensuremath{\Delta}\xspace}                 
 \def\PXi      {\ensuremath{\Xi}\xspace}                 
 \def\PLambda      {\ensuremath{\Lambda}\xspace}                 
 \def\PSigma      {\ensuremath{\Sigma}\xspace}                 
 \def\POmega      {\ensuremath{\Omega}\xspace}                 
 \def\PUpsilon      {\ensuremath{\Upsilon}\xspace}                 
 
 \def\PB      {\ensuremath{\mathrm{B}}\xspace}                 
                  
 \def\PD      {\ensuremath{\mathrm{D}}\xspace}

 \def\PJ      {\ensuremath{\mathrm{J}}\xspace}                 
 \def\PK      {\ensuremath{\mathrm{K}}\xspace}

 \def\PS      {\ensuremath{\mathrm{S}}\xspace}

 \def\Pb      {\ensuremath{\mathrm{b}}\xspace}                 
 \def\Pc      {\ensuremath{\mathrm{c}}\xspace}                 
 \def\Pd      {\ensuremath{\mathrm{d}}\xspace}                 
 \def\Pe      {\ensuremath{\mathrm{e}}\xspace}

 \def\Pi      {\ensuremath{\mathrm{i}}\xspace}

 \def\Pp      {\ensuremath{\mathrm{p}}\xspace}

 \def\Ps      {\ensuremath{\mathrm{s}}\xspace}

}
{

 \def\Pmu         {\ensuremath{\mu}\xspace}

 \def\Ppi         {\ensuremath{\pi}\xspace}

 \def\Ppsi        {\ensuremath{\psi}\xspace}                 
                  
 \mathchardef\PDelta="7101
 \mathchardef\PXi="7104
 \mathchardef\PLambda="7103
 \mathchardef\PSigma="7106
 \mathchardef\POmega="710A
 \mathchardef\PUpsilon="7107
                  
 \def\PB      {\ensuremath{B}\xspace}                 
                  
 \def\PD      {\ensuremath{D}\xspace}

 \def\PJ      {\ensuremath{J}\xspace}                 
 \def\PK      {\ensuremath{K}\xspace}

 \def\PS      {\ensuremath{S}\xspace}

 \def\Pb      {\ensuremath{b}\xspace}                 
 \def\Pc      {\ensuremath{c}\xspace}                 
 \def\Pd      {\ensuremath{d}\xspace}                 
 \def\Pe      {\ensuremath{e}\xspace}

 \def\Pi      {\ensuremath{i}\xspace}

 \def\Pp      {\ensuremath{p}\xspace}

 \def\Ps      {\ensuremath{s}\xspace}

}

\makeatletter
\ifcase \@ptsize \relax% 10pt
  \newcommand{\miniscule}{\@setfontsize\miniscule{4}{5}}% \tiny: 5/6
\or% 11pt
  \newcommand{\miniscule}{\@setfontsize\miniscule{5}{6}}% \tiny: 6/7
\or% 12pt
  \newcommand{\miniscule}{\@setfontsize\miniscule{5}{6}}% \tiny: 6/7
\fi
\makeatother

\DeclareRobustCommand{\optbar}[1]{\shortstack{{\miniscule (\rule[.5ex]{1.25em}{.18mm})}
  \\ [-.7ex] $#1$}}

\def\en         {{\ensuremath{\Pe^-}}\xspace}   % electron negative (\em is taken)
\def\ep         {{\ensuremath{\Pe^+}}\xspace}

\def\mup        {{\ensuremath{\Pmu^+}}\xspace}
\def\mun        {{\ensuremath{\Pmu^-}}\xspace} % muon negative (\mum is taken)
\def\mumu       {{\ensuremath{\Pmu^+\Pmu^-}}\xspace}

\def\dquark    {{\ensuremath{\Pd}}\xspace}

\def\squark    {{\ensuremath{\Ps}}\xspace}

\def\cquark    {{\ensuremath{\Pc}}\xspace}

\def\bquark    {{\ensuremath{\Pb}}\xspace}

\def\pion   {{\ensuremath{\Ppi}}\xspace}

\def\pip    {{\ensuremath{\pion^+}}\xspace}
\def\pim    {{\ensuremath{\pion^-}}\xspace}

\def\kaon    {{\ensuremath{\PK}}\xspace}
  \def\Kbar    {{\kern 0.2em\overline{\kern -0.2em \PK}{}}\xspace}

\def\KorKbar    {\kern 0.18em\optbar{\kern -0.18em K}{}\xspace}

\def\Kp      {{\ensuremath{\kaon^+}}\xspace}
\def\Km      {{\ensuremath{\kaon^-}}\xspace}

  \def\Dbar    {{\kern 0.2em\overline{\kern -0.2em \PD}{}}\xspace}
\def\D       {{\ensuremath{\PD}}\xspace}

\def\DorDbar    {\kern 0.18em\optbar{\kern -0.18em D}{}\xspace}
\def\Dz      {{\ensuremath{\D^0}}\xspace}

\def\Dm      {{\ensuremath{\D^-}}\xspace}

\def\Dstarp  {{\ensuremath{\D^{*+}}}\xspace}

\def\Dsm     {{\ensuremath{\D^-_\squark}}\xspace}

\def\B       {{\ensuremath{\PB}}\xspace}
\def\Bbar    {{\ensuremath{\kern 0.18em\overline{\kern -0.18em \PB}{}}}\xspace}

\def\BorBbar    {\kern 0.18em\optbar{\kern -0.18em B}{}\xspace}
\def\Bz      {{\ensuremath{\B^0}}\xspace}

\def\Bu      {{\ensuremath{\B^+}}\xspace}

\def\Bp      {{\ensuremath{\Bu}}\xspace}

\def\Bd      {{\ensuremath{\B^0}}\xspace}
\def\Bs      {{\ensuremath{\B^0_\squark}}\xspace}

\def\jpsi     {{\ensuremath{{\PJ\mskip -3mu/\mskip -2mu\Ppsi\mskip 2mu}}}\xspace}
\def\psitwos  {{\ensuremath{\Ppsi{(2\PS)}}}\xspace}

  \def\Y#1S{\ensuremath{\PUpsilon{(#1S)}}\xspace}% no space before {...}!

\def\proton      {{\ensuremath{\Pp}}\xspace}
\def\antiproton  {{\ensuremath{\overline \proton}}\xspace}

\def\Lz          {{\ensuremath{\PLambda}}\xspace}
\def\Lbar        {{\ensuremath{\kern 0.1em\overline{\kern -0.1em\PLambda}}}\xspace}
\def\LorLbar    {\kern 0.18em\optbar{\kern -0.18em \PLambda}{}\xspace}

\def\Lb      {{\ensuremath{\Lz^0_\bquark}}\xspace}
\def\Lbbar   {{\ensuremath{\Lbar{}^0_\bquark}}\xspace}
\def\Lc      {{\ensuremath{\Lz^+_\cquark}}\xspace}

\def\BF         {{\ensuremath{\mathcal{B}}}\xspace}

\def\BR         {\BF}
\newcommand{\decay}[2]{\ensuremath{#1\!\to #2}\xspace}         % {\Pa}{\Pb \Pc}

\def\to                 {\ensuremath{\rightarrow}\xspace}

\def\jpsimm {{\decay{\jpsi}{\mup\mun}\xspace}}
\def\psimm {{\decay{\psitwos}{\mup\mun}\xspace}}
\def\psijpp {{\decay{\psitwos}{\jpsi\pip\pim}\xspace}}
\def\jpp {{{\jpsi\pip\pim}\xspace}}
\def\jpipiKp {{\jpsi\pip\pim\proton\Km\xspace}}
\def\LbJpsiPK {{\decay{\Lb}{\jpsi\proton\Km}\xspace}}

\def\LbPsiPK {{\decay{\Lb}{\psitwos\proton\Km}\xspace}}

\def\LbPsimmPK {{\decay{\Lb}{\psitwos[\to\mumu]\proton\Km}\xspace}}
\def\LbPsijppPK {{\decay{\Lb}{\psitwos[\to\jpp]\proton\Km}\xspace}}
\def\LbJpipiPK {{\decay{\Lb}{\jpipiKp}\xspace}}

\def\AT#1     {\ensuremath{A_{\mathrm{T}}^{#1}}\xspace}           % 2

\def\C#1      {\ensuremath{\mathcal{C}_{#1}}\xspace}                       % 9
\def\Cp#1     {\ensuremath{\mathcal{C}_{#1}^{'}}\xspace}                    % 7
\def\Ceff#1   {\ensuremath{\mathcal{C}_{#1}^{\mathrm{(eff)}}}\xspace}        % 9  
\def\Cpeff#1  {\ensuremath{\mathcal{C}_{#1}^{'\mathrm{(eff)}}}\xspace}       % 7
\def\Ope#1    {\ensuremath{\mathcal{O}_{#1}}\xspace}                       % 2
\def\Opep#1   {\ensuremath{\mathcal{O}_{#1}^{'}}\xspace}                    % 7

\newcommand{\tev}{\ifthenelse{\boolean{inbibliography}}{\ensuremath{~T\kern -0.05em eV}\xspace}{\ensuremath{\mathrm{\,Te\kern -0.1em V}}}\xspace}
\newcommand{\gev}{\ensuremath{\mathrm{\,Ge\kern -0.1em V}}\xspace}
\newcommand{\mev}{\ensuremath{\mathrm{\,Me\kern -0.1em V}}\xspace}
\newcommand{\kev}{\ensuremath{\mathrm{\,ke\kern -0.1em V}}\xspace}
\newcommand{\ev}{\ensuremath{\mathrm{\,e\kern -0.1em V}}\xspace}
\newcommand{\gevc}{\ensuremath{{\mathrm{\,Ge\kern -0.1em V\!/}c}}\xspace}
\newcommand{\mevc}{\ensuremath{{\mathrm{\,Me\kern -0.1em V\!/}c}}\xspace}
\newcommand{\gevcc}{\ensuremath{{\mathrm{\,Ge\kern -0.1em V\!/}c^2}}\xspace}
\newcommand{\gevgevcccc}{\ensuremath{{\mathrm{\,Ge\kern -0.1em V^2\!/}c^4}}\xspace}
\newcommand{\mevcc}{\ensuremath{{\mathrm{\,Me\kern -0.1em V\!/}c^2}}\xspace}

\def\mum  {\ensuremath{{\,\upmu\mathrm{m}}}\xspace}

\def\invfb   {\ensuremath{\mbox{\,fb}^{-1}}\xspace}

\def\ps   {\ensuremath{{\mathrm{ \,ps}}}\xspace}

\newcommand{\chisq}{\ensuremath{\chi^2}\xspace}

\def\gsim{{~\raise.15em\hbox{$>$}\kern-.85em
          \lower.35em\hbox{$\sim$}~}\xspace}
\def\lsim{{~\raise.15em\hbox{$<$}\kern-.85em
          \lower.35em\hbox{$\sim$}~}\xspace}

\def\sPlot{\mbox{\em sPlot}\xspace}
\def\sqs   {\ensuremath{\protect\sqrt{s}}\xspace}

\def\ptot       {\mbox{$p$}\xspace}
\def\pt         {\mbox{$p_{\mathrm{ T}}$}\xspace}

\def\evtgen     {\mbox{\textsc{EvtGen}}\xspace}

\def\geant      {\mbox{\textsc{Geant4}}\xspace}

\def\photos     {\mbox{\textsc{Photos}}\xspace}

\def\pythia     {\mbox{\textsc{Pythia}}\xspace}

\def\tell1  {TELL1\xspace}
\def\ukl1   {UKL1\xspace}

\newcommand{\eg}{\mbox{\itshape e.g.}\xspace}
\newcommand{\ie}{\mbox{\itshape i.e.}\xspace}

\usepackage{cite} % Allows for ranges in citations
\usepackage{mciteplus}

\usepackage{multirow}
\usepackage{rotating}

\usepackage{comment}

\usepackage{cancel}

\begin{document}

%%%%%%%%%%%%%%%%%%%%%%%%%
%%%%% Title     %%%%%%%%%
%%%%%%%%%%%%%%%%%%%%%%%%%
\renewcommand{\thefootnote}{\fnsymbol{footnote}}
\setcounter{footnote}{1}

% %%%%%%% CHOOSE TITLE PAGE--------
%\onecolumn
%\input{title-LHCb-INT}
%\input{title-LHCb-ANA}
%\input{title-LHCb-CONF}
% $Id: title-LHCb-PAPER.tex 92924 2016-06-01 15:22:52Z ibelyaev $
% ===============================================================================
% Purpose: LHCb-PAPER journal paper title page template
% Author: 
% Created on: 2010-09-25
% ===============================================================================

%%%%%%%%%%%%%%%%%%%%%%%%%
%%%%%  TITLE PAGE  %%%%%%
%%%%%%%%%%%%%%%%%%%%%%%%%
\begin{titlepage}
\pagenumbering{roman}

% Header ---------------------------------------------------
\vspace*{-1.5cm}
\centerline{\large EUROPEAN ORGANIZATION FOR NUCLEAR RESEARCH (CERN)}
\vspace*{1.5cm}
\noindent
\begin{tabular*}{\linewidth}{lc@{\extracolsep{\fill}}r@{\extracolsep{0pt}}}
\ifthenelse{\boolean{pdflatex}}% Logo format choice
{\vspace*{-2.7cm}\mbox{\!\!\!\includegraphics[width=.14\textwidth]{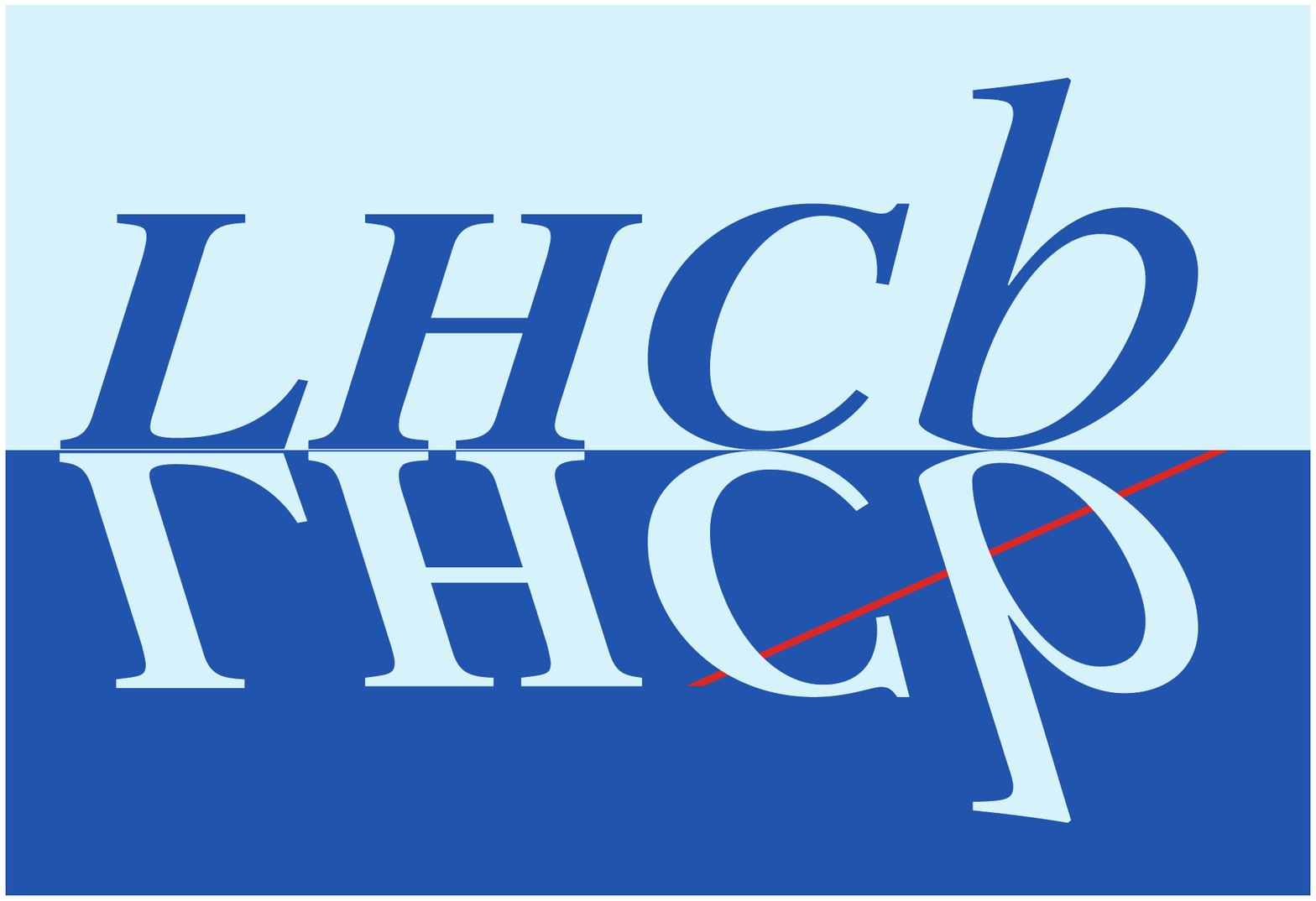}} & &}%
{\vspace*{-1.2cm}\mbox{\!\!\!\includegraphics[width=.12\textwidth]{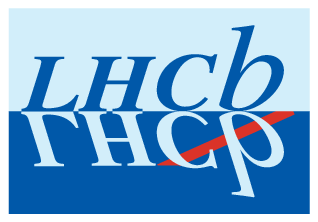}} & &}%
\\
 & & CERN-EP-2016-070 \\  % ID
 & & LHCb-PAPER-2015-060 \\  % ID
%% & & \today \\ % Date - Can also hardwire e.g.: 23 March 2010
 & & March 24, 2016 \\ % Date - Can also hardwire e.g.: 23 March 2010
 & & \\
% not in paper \hline
\end{tabular*}

%\vspace*{4.0cm}
\vspace*{1cm}

% Title --------------------------------------------------
{\normalfont\bfseries\boldmath\huge
\begin{center}
  Observation of
  ${\LbPsiPK}$ and
  \mbox{$\Lb\to\jpsi\pip\pim\proton\Km$} 
  decays and a measurement of  the \Lb~baryon~mass
\end{center}
}

%\vspace*{2.0cm}
\vspace*{0.5cm}

% Authors -------------------------------------------------
\begin{center}
%In the footnote, replace 'paper' by 'letter' in case of submission to PRL or PLB 
The LHCb collaboration\footnote{Authors are listed at the end of this paper.}
\end{center}

\vspace{\fill}

% Abstract -----------------------------------------------
\begin{abstract}
  \noindent
  The~decays $\LbPsiPK$ and $\LbJpipiPK$ are observed in a~data sample corresponding to
  an~integrated luminosity of $3\invfb$, 
  collected 
  in proton-proton collisions at 7 and 8\tev centre-of-mass energies
  by the~\lhcb detector.
  The \psitwos mesons are reconstructed through the decay modes $\psimm$ and $\psijpp$.
  The~branching fractions relative to that of $\LbJpsiPK$ are measured to be 
  \begin{align*}
    \dfrac{\BR(\LbPsiPK)}{\BR(\LbJpsiPK)}   &= (20.70\pm 0.76\pm 0.46\pm 0.37)\times10^{-2} \,,\\
    \dfrac{\BR(\LbJpipiPK)}{\BR(\LbJpsiPK)} &= (20.86\pm 0.96\pm 1.34)\times10^{-2} \,,
  \end{align*}
  where the~first uncertainties are statistical, the~second are systematic and the third is related to
  the~knowledge of \jpsi and \psitwos branching fractions.
  The~mass of the~\Lb~baryon is measured to be 
  \begin{equation*}
    M(\Lb) = 5619.65 \pm 0.17 \pm 0.17\mevcc\,, %,
  \end{equation*}
  where the~uncertainties are statistical and systematic.
\end{abstract}

\vspace*{0.3cm}

\begin{center}
  Published in JHEP 1605 (2016) 132
\end{center}

\vspace{\fill}

{\footnotesize 
\centerline{\copyright~CERN on behalf of the \lhcb collaboration, licence \href{http://creativecommons.org/licenses/by/4.0/}{CC-BY-4.0}.}}
\vspace*{2mm}

\end{titlepage}

%%%%%%%%%%%%%%%%%%%%%%%%%%%%%%%%
%%%%%  EOD OF TITLE PAGE  %%%%%%
%%%%%%%%%%%%%%%%%%%%%%%%%%%%%%%%

%  empty page follows the title page ----
\newpage
\setcounter{page}{2}
\mbox{~}

%\newpage
%
%% Author List ----------------------------
%%  You need to get a new author list!
%\input{LHCb_authorlist.tex}
%
%The author list for journal publications is provided by the Membership Committee shortly after 'approval to go to paper' has been given.
%%It will be made available on the page
%%\verb!http://www.physik.uzh.ch/~strauman/forMemCo/LHCb-PAPER-XXXX-XXX/! .
%It will be sent to you by email shortly after a paper number has beens assigned.
%The author list should be included already at first circulation, 
%to allow new members of the collaboration to verify whether they have been included correctly.
%Occasionally a misspelled name is corrected or associated institutions become full members.
%In that case, a new author list will be sent to you.
%In case line numbering doesn't work well after including the authorlist, try moving the \verb!\bigskip! after the last author to a separate line.
%
%
%The authorship for Conference Reports should be ``The LHCb
%  collaboration'', with a footnote giving the name(s) of the contact
%  author(s), but without the full list of collaboration names.

\cleardoublepage

%\twocolumn
% %%%%%%%%%%%%% ---------

\renewcommand{\thefootnote}{\arabic{footnote}}
\setcounter{footnote}{0}

%%%%%%%%%%%%%%%%%%%%%%%%%%%%%%%%
%%%%%  Table of Content   %%%%%%
%%%%%%%%%%%%%%%%%%%%%%%%%%%%%%%%
%%%% Uncomment next 2 lines if desired
%\tableofcontents
%\cleardoublepage

%%%%%%%%%%%%%%%%%%%%%%%%%
%%%%% Main text %%%%%%%%%
%%%%%%%%%%%%%%%%%%%%%%%%%

\pagestyle{plain} % restore page numbers for the main text
\setcounter{page}{1}
\pagenumbering{arabic}

%% Uncomment during review phase. 
%% Comment before a final submission.
%%  \linenumbers

% You can include short sections directly in the main tex file.
% However, for larger papers it is desirable to split the text into
% several semiautonomous files, which can be revised independently.
% This is especially useful when developing a document in
% collaboration with several people, since then different parts can be
% edited independently.  This type of file organization is shown here.

% $Id: introduction.tex 89652 2016-03-22 09:14:49Z ibelyaev $

\section{Introduction}
\label{sec:Introduction}
The \Lb baryon is the~isospin singlet ground state of a~bottom quark and two light quarks.
The~rich phenomenology associated with decays of bottom baryons
allows many measurements of masses, lifetimes 
and branching fractions, which test the~theoretical understanding 
of weak decays of heavy hadrons in the~framework of heavy quark 
effective theory\,(HQET) and the~underlying QCD physics~\cite{Neubert:1993mb,*Manohar:2000dt}.
At~the~Tevatron,
properties of the~\Lb baryon, such as mass and lifetime, have been measured using two\nobreakdash-body modes, specifically
$\decay{\Lb}{\jpsi\Lz^0}$ and
$\decay{\Lb}{\Lc\pim}$ decays~\cite{Abulencia:2006df,Abazov:2011wt,LbMassCDF}.\footnote{The~inclusion
  of charge\nobreakdash-conjugate modes is implied throughout this paper.}
The~high production rate of \bquark~quarks at the~Large Hadron Collider\,(LHC), along with
the~excellent momentum and mass resolution and
the~hadron identification capabilities of the~LHCb detector, open up a~host of multibody
and Cabibbo-suppressed decay channels of \Lb~baryons, \eg the~decays
\mbox{$\Lb\to\Dz\proton\Km$},
\mbox{$\Lb\to\Lc\Km$}~\cite{LHCb-PAPER-2013-056},
\mbox{$\Lb\to\Lc\Dm$},
\mbox{$\Lb\to\Lc\Dsm$}~\cite{LHCb-PAPER-2014-002} 
and
\mbox{$\Lb\to\jpsi\proton\pim$}~\cite{LHCb-PAPER-2014-020}.
The~high signal yield of the \mbox{$\Lb\to\jpsi\proton\Km$}~decay~\cite{LHCb-PAPER-2015-032}
allowed the precise measurement of the \Lb lifetime~\mbox{\cite{LHCb-PAPER-2014-003,LHCb-PAPER-2014-048}}.
The recent analysis of this decay mode uncovered a double resonant structure in the~$\jpsi\proton$~system
consistent with two 
pentaquark states~\cite{LHCb-PAPER-2015-029}.
\lhcb has also measured several \B~meson decays into final states with
charmonia~\cite{LHCb-PAPER-2012-010,LHCb-PAPER-2012-053,LHCb-PAPER-2013-024,LHCb-PAPER-2014-050,LHCb-PAPER-2014-056,LHCb-PAPER-2015-024}.
The~first observation of $\Lb$ decays to excited charmonium, the $\Lb\to\psitwos\PLambda^0$ decay, has been presented by the \atlas 
collaboration~\cite{AtlasLb}.
An~experimental investigation of other similar  multibody decays of the~\Lb~baryon
should lead to deeper insights into QCD.

In this paper, the~first observations of the~decays~$\LbPsiPK$ and
$\LbJpipiPK$ are reported,
where $\psitwos$~mesons are reconstructed in
the~final states $\mumu$~and $\jpsi\pip\pim$.
The~ratios of the~branching fractions of these decays to that of
the normalization decay $\Lb\to\jpsi\proton\Km$,
%%\begin{subequations}
%%  \label{eq:rdef}
  \begin{eqnarray}
    R^{\psitwos}     & \equiv &  \dfrac{\BR(\LbPsiPK)}{\BR(\LbJpsiPK)} \label{eq:rdef1} \, , \\ 
    R^{\jpsi\pip\pim} & \equiv &  \dfrac{\BR(\LbJpipiPK)}{\BR(\LbJpsiPK)} \label{eq:rdef2} \, ,
  \end{eqnarray}
%%\end{subequations}
are measured.
In~measuring the~branching fraction of $\LbJpipiPK$~decays,  contributions via intermediate resonances,
such as $\LbPsiPK$, are implicitly included.
The~low energy release in these decays allows a precise determination of the~\Lb mass with
a~small systematic uncertainty.

This study is based on a~data sample corresponding to an integrated luminosity of~3\invfb,
collected with the~\lhcb detector in~$\proton\proton$~collisions at
centre\nobreakdash-of\nobreakdash-mass energies $\sqs=7$~and~$8\tev$.

\section{Detector and simulation}
\label{sec:Detector}

The \lhcb detector~\cite{Alves:2008zz,LHCb-DP-2014-002} is a single-arm forward
spectrometer covering the \mbox{pseudorapidity} range $2<\eta <5$,
designed for the study of particles containing \bquark or \cquark
quarks. The detector includes a high-precision tracking system
consisting of a silicon-strip vertex detector surrounding the $\proton\proton$
interaction region, a large-area silicon-strip detector located
upstream of a dipole magnet with a bending power of about 
$4{\mathrm{\,Tm}}$, and three stations of silicon-strip detectors and straw drift tubes placed downstream of the magnet.
The~polarity of the~dipole magnet is reversed periodically throughout data-taking.
The~tracking system provides a~measurement of the~momentum, \ptot, of charged particles with
a relative uncertainty that varies from 0.5\% at low momentum to 1.0\% at 200\gevc.
The minimum distance of a~track to a~primary vertex\,(PV), the~impact parameter, is measured with
a~resolution of $(15+29/\pt)\mum$,
where \pt is the component of the momentum transverse to the beam, in\,\gevc~\cite{LHCb-DP-2014-001}.
Large samples of $\Bp\to\jpsi\Kp$ and $\jpsi\to\mup\mun$ decays, 
collected concurrently with the data set, were used to calibrate the momentum scale of the spectrometer to a~precision of~$0.03\,\%$
\cite{LHCb-PAPER-2012-048}. 

Different types of charged hadrons are distinguished using information from two ring-imaging Cherenkov detectors (RICH). 
Photons, electrons and hadrons are identified by a~calorimeter system consisting of scintillating-pad and preshower detectors, an electromagnetic
calorimeter and a hadronic calorimeter. Muons are identified by a system composed of alternating layers of iron and multiwire
proportional chambers.

The trigger~\cite{LHCb-DP-2012-004} comprises two stages.
Events are first required to pass the~hardware trigger, 
which selects
muon candidates with $\pt>1.48\,(1.76)\gevc$ or
pairs of opposite-sign muon candidates 
with a~requirement that the~product of the~muon transverse momenta 
is larger than $1.7\,(2.6)\,\mathrm{GeV}^2/c^2$ 
%%$\sqrt{p^{\mup}_{\mathrm{T}}\times p^{\mun}_{\mathrm{T}}}>1.3\,(1.6)\gevc$ 
for data collected at $\sqrt{s}=7\,(8)\tev$.
The subsequent software trigger is composed of two stages, 
the~first of which performs a~partial event reconstruction,
while full event reconstruction is done at the~second stage.
At the~first stage of the~software trigger 
the invariant mass of well\nobreakdash-reconstructed pairs of 
oppositely charged muons forming a~good\nobreakdash-quality
two\nobreakdash-track vertex 
is required to exceed 2.7\gevcc,
and the~two\nobreakdash-track vertex 
is required to be significantly displaced from all PVs.

The~analysis technique reported below has been validated using simulated events. 
The~$\proton\proton$~collisions are generated using \pythia~\cite{Sjostrand:2006za,*Sjostrand:2007gs} with a specific \lhcb
configuration~\cite{LHCb-PROC-2010-056}. Decays of hadronic particles are described by \evtgen~\cite{Lange:2001uf}, in which final-state
radiation is generated using \photos~\cite{Golonka:2005pn}. The interaction of the generated particles with the~detector, and its response,
are implemented using the \geant toolkit~\cite{Allison:2006ve, *Agostinelli:2002hh} as described in~Ref.~\cite{LHCb-PROC-2011-006}.

\section{Event selection}
\label{sec:Selection}

The decays
$\LbPsiPK$,
$\LbJpipiPK$ and 
$\LbJpsiPK$
are reconstructed using decay modes $\psimm$, $\psitwos\to\jpsi\pip\pim$ 
and $\jpsimm$.
Common selection criteria, based on those
used in Refs.~\cite{LHCb-PAPER-2014-009,LHCb-PAPER-2014-056}, are
used for all channels,  except for those related to the~selection
of two additional pions in the~$\LbJpipiPK$ and $\Lb\to\psitwos[\to\jpp]\proton\Km$
channels.

Muon, proton, kaon and pion candidates are selected from 
well\nobreakdash-reconstructed tracks within the acceptance of the
spectrometer that are identified using information from
the~RICH, calorimeter and muon detectors~\cite{LHCb-DP-2012-003,LHCb-DP-2013-001}. 
Muons, protons, kaons and pions are required to have a transverse momentum larger than $550$,
$800$, $500$ and $200\mevc$,  respectively.
To~allow good particle identification,
kaons and pions are required to~have a~momentum between $3.2 \gevc$ and $150 \gevc$ whilst protons must have
a~momentum between $10 \gevc$ and $150 \gevc$.
To~reduce combinatorial background involving tracks from the primary
$\proton\proton$~interaction vertices, only tracks that
exceed a~minimum impact parameter \chisq with respect to every PV are used.
The~impact parameter \chisq is defined as the~difference between the~\chisq of the~PV
reconstructed with and without the~considered particle.

Pairs of oppositely\nobreakdash-charged muons
originating from a~common vertex are combined to form $\jpsimm$~or 
$\psimm$~candidates. The~resulting dimuon candidates are required to 
have an~invariant mass between
$-5\sigma$ and $+3\sigma$ around the known $\jpsi$~or $\psitwos$~masses~\cite{PDG2014}, where $\sigma$ 
is the~mass resolution.
An~asymmetric mass interval is chosen to include part of the low-mass tail due to final-state radiation.

Candidate \Lb~baryons are formed from $\jpsi\proton\Km$, $\psitwos\proton\Km$ and $\jpp\proton\Km$ combinations.
Each~candidate is associated with the~PV with respect to which
it has the~smallest impact parameter significance.
The~\Lb~mass resolution is improved  by employing a~kinematic fit~\cite{Hulsbergen:2005pu}
that  constrains the~mass of the \jpsi~candidate to its known value and 
requires the momentum of the \Lb~candidate to point back to the~PV. 
A~requirement on the quality of this fit is applied to further suppress combinatorial background.
Finally, the measured decay time of the~\Lb~candidate, calculated with respect to the~associated primary vertex, 
is required to be between 
0.5 and 6.7\ps.
The lower limit is used to suppress background from particles coming from the~PV
while the upper limit removes poorly reconstructed candidates.

To~suppress cross\nobreakdash-feed from 
decays of the \Bs~meson into 
\mbox{$\jpsi\Km\Kp$},
\mbox{$\psitwos\Km\Kp$} and 
\mbox{$\jpsi\pip\pim\Km\Kp$}~final states, 
with the positively\nobreakdash-charged kaon
misidentified as a proton, 
a~veto on the~\Lb~candidate mass,
recalculated with a~kaon mass hypothesis for the~proton,
is applied.
Any~candidate~with a~recalculated mass 
consistent with the nominal \Bs~mass is rejected.
%%within $\pm3\sigma$ of the~nominal
%%\Bs~mass is rejected, where $\sigma$ is mass resolution resolution.
A~similar veto is applied to suppress
cross\nobreakdash-feed from
decays of \Bd~mesons into 
\mbox{$\jpsi\Km\pip$},
\mbox{$\psitwos\Km\pip$} and 
\mbox{$\jpsi\pim\pip\pip\Km$}~decays with the positively\nobreakdash-charged pion
misidentified as a~proton.

% \section{Signal yields}	% v0.1
\section{Measurement of branching fractions}	% Vanya's proposal
\label{sec:signals}

\subsection{Signal yields and efficiencies}\label{sec:signalpeaks}

\begin{figure}[t]
  \setlength{\unitlength}{1mm}
  \centering
  \begin{picture}(150,67)
    \put( 0,2){\includegraphics*[width=75mm,height=65mm]{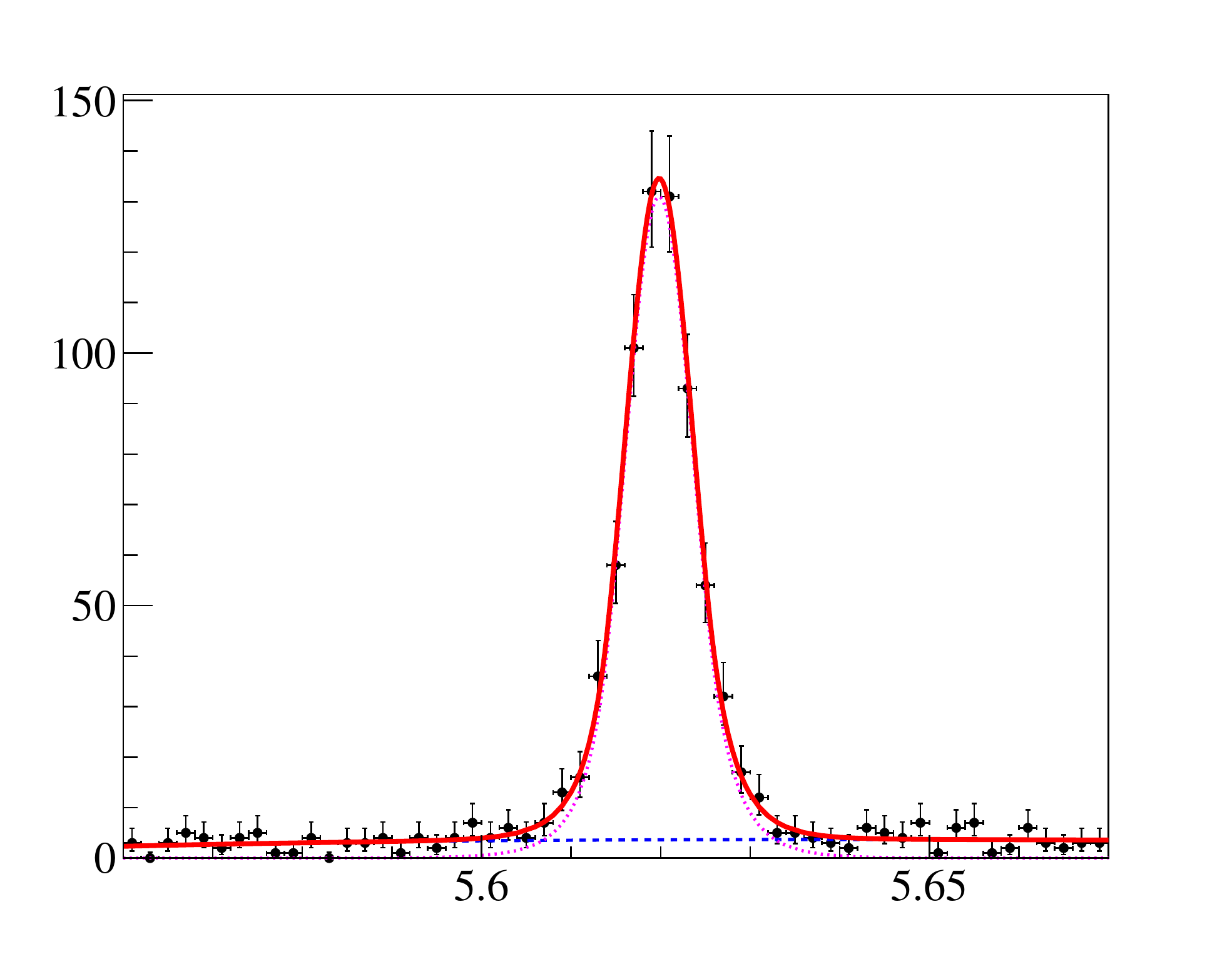}}
    \put(80,2){\includegraphics*[width=75mm,height=65mm]{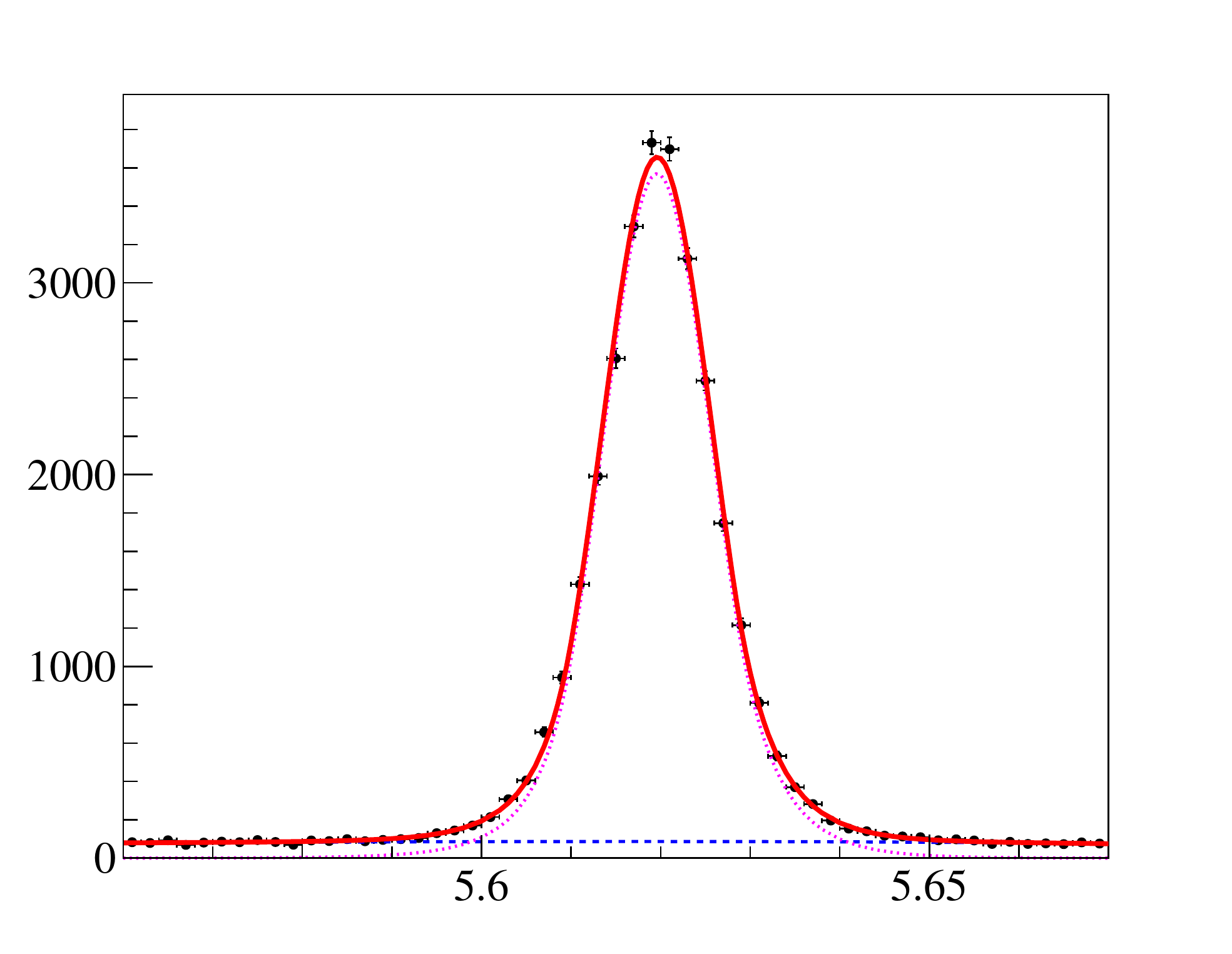}}
    \put( -5,20 )  { \begin{sideways}Candidates/(2$\mevcc$)\end{sideways}  }
    \put( 75,20 )  { \begin{sideways}Candidates/(1$\mevcc$)\end{sideways}  }
    \put( 105,0 )  { $ m(\jpsi\proton\Km)$ }
    \put(   6,0 )  { $ m(\psitwos[\to\mumu]\proton\Km)$ }
    \put( 51,0  )  { $\left[\gevcc\right]$ }
    \put( 131,0 )  { $\left[\gevcc\right]$ }
    \put( 10,53  ){ \small LHCb}
    \put( 90,53  ){ \small LHCb}
  \end{picture}
  \caption { \small Mass distributions of selected
    (left)~$\LbPsimmPK$
    and
    (right)~$\LbJpsiPK$~candidates.
    The~total fit 
    function\,(solid red),
    the~$\Lb$ signal contribution\,(dotted magenta)
    and the~combinatorial background\,(dashed blue) are shown.
    The~error bars show 68\%~Poissonian confidence intervals.}
  \label{fig:Fig_psimm}
\end{figure}

The~mass distributions for selected $\LbPsimmPK$~candidates
and candidates for the~normalization channel $\LbJpsiPK$ are shown in Fig.~\ref{fig:Fig_psimm}.
Signal yields are determined using unbinned extended maximum likelihood fits to these distributions.
The~signal is modelled with a~modified Gaussian function with power\nobreakdash-law tails
on~both sides~\cite{Skwarnicki:1986xj,LHCb-PAPER-2011-013}, 
where the~tail parameters are fixed from~simulation and the mass resolution parameter is allowed to vary.
The~background is modelled with an~exponential function multiplied 
by a~first\nobreakdash-order polynomial.
The resolution parameters obtained from the~fits are found to be \mbox{$3.82\pm0.17\mevcc$} for
the~channel $\LbPsimmPK$
and \mbox{$6.12\pm0.05\mevcc$} for $\LbJpsiPK$, in good agreement with expectations from simulation.

\begin{figure}[t]
  \setlength{\unitlength}{1mm}
  \centering
  \begin{picture}(150,67)
    \put(0,2){\includegraphics*[width=75mm,height=65mm]{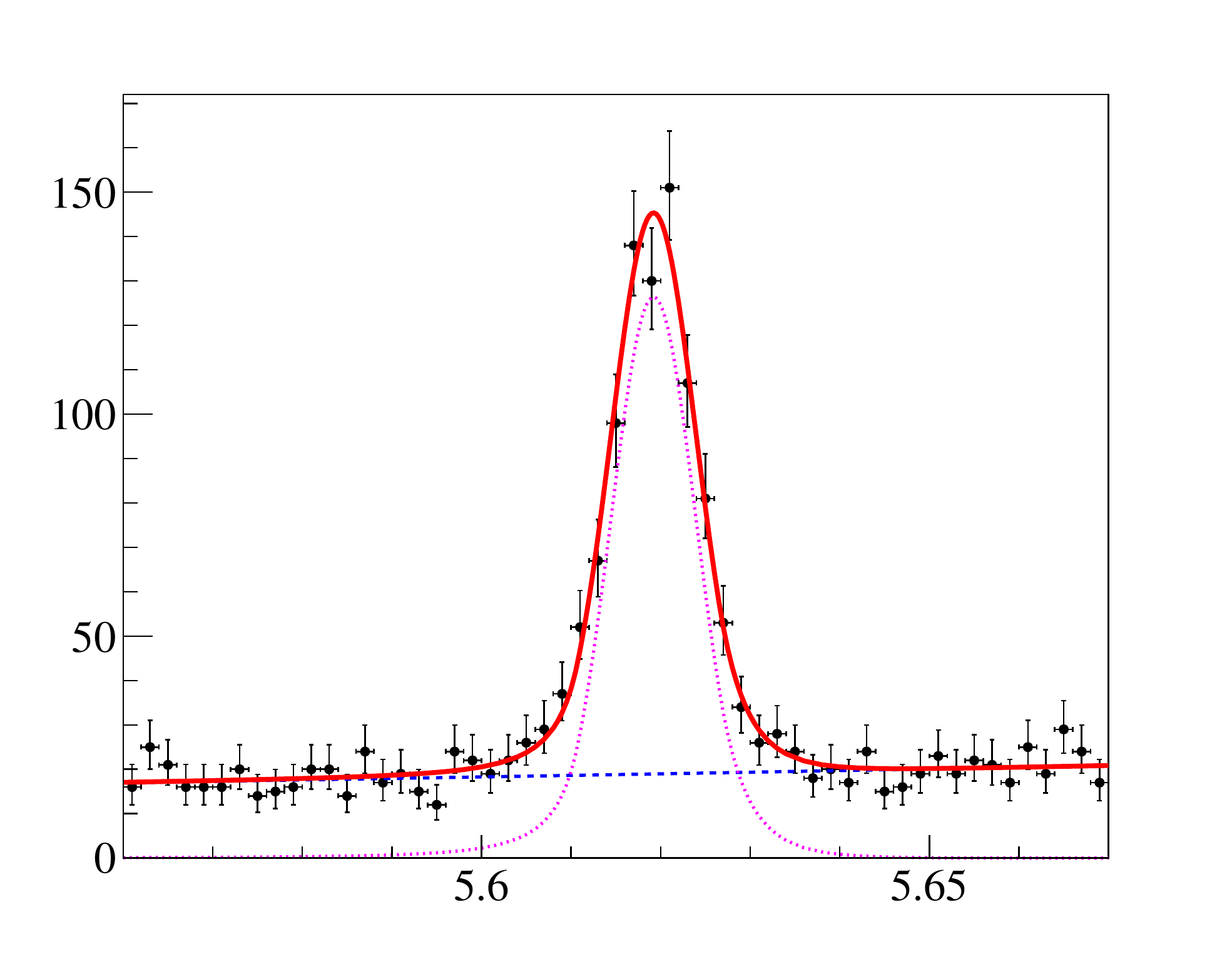}}
    \put( -5,20 )  { \begin{sideways}Candidates/(2$\mevcc$)\end{sideways}  }
    \put( 15,0  )  { $m(\jpp\proton\Km)$ }
    \put(80,2){\includegraphics*[width=75mm,height=65mm]{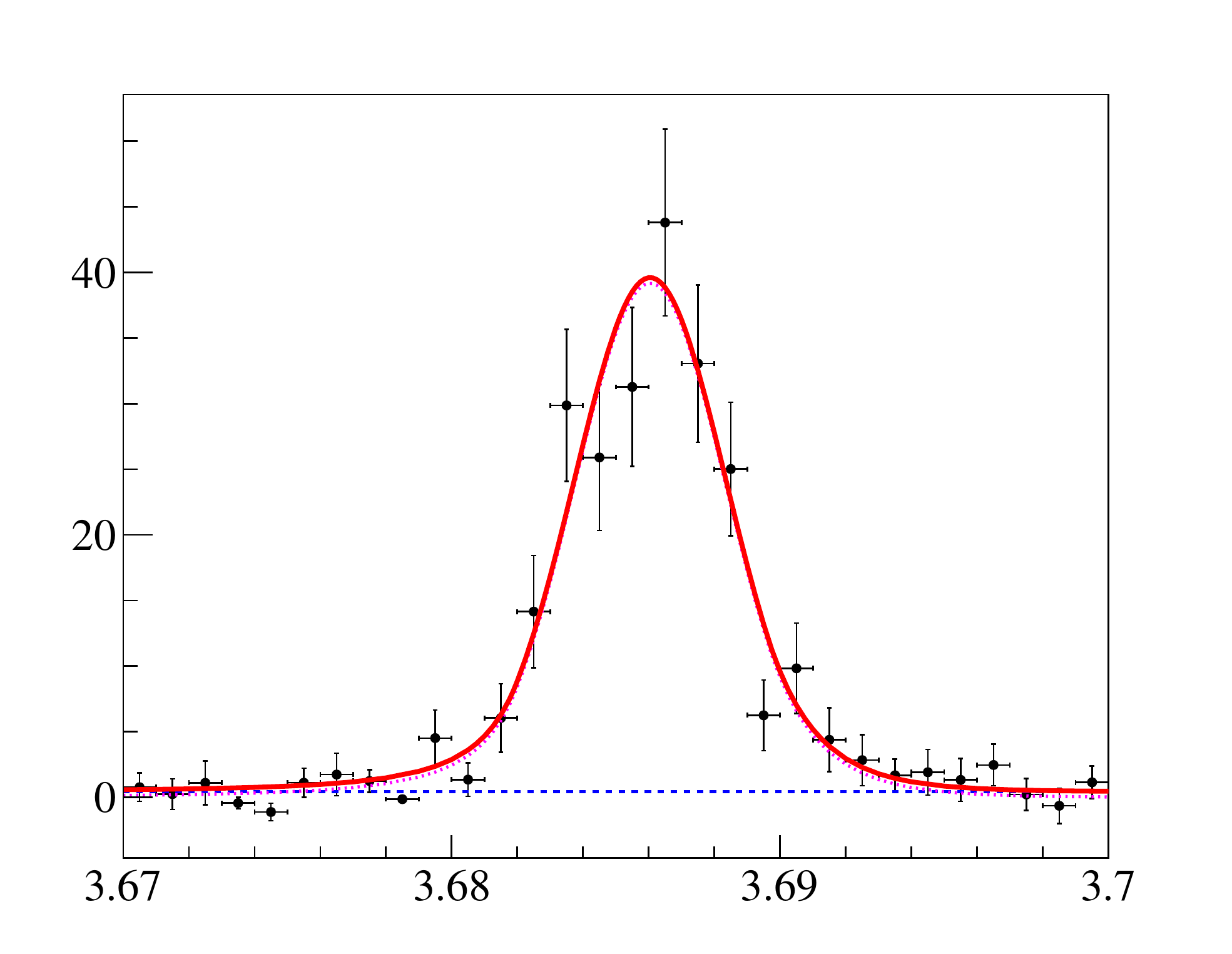}}
    \put( 75,20  )  { \begin{sideways}Candidates/(1$\mevcc$)\end{sideways}  }
    \put( 102,0  )  { $m(\jpp)$ }
    \put( 51,0  )  { $\left[\gevcc\right]$ }
    \put( 131,0 )  { $\left[\gevcc\right]$ }
    \put( 10,53  ){ \small LHCb}
    \put( 90,53  ){ \small LHCb}
  \end{picture}
  \caption { \small (left)~Mass distribution of selected $\LbJpipiPK$ candidates. 
    (right)~Background-subtracted $\jpp$ mass distribution for that mode.
    The total fit function and the signal contributions are shown by solid red and dotted magenta lines, respectively. The combinatorial background in the left plot 
    and nonresonant contribution in the right plot are shown by dashed blue lines. 
  }
  \label{fig:Fig_jpipi}
\end{figure}

The mass distribution for selected $\LbJpipiPK$ candidates is shown in Fig.~\ref{fig:Fig_jpipi}(left),
along with the~result of 
an~unbinned extended maximum likelihood fit using the~model described above.
The~mass resolution parameter obtained from the~fit is \mbox{$4.72\pm0.23\mevcc$}.
The~mass distribution of the~\jpp~system from signal $\LbJpipiPK$ decays
is presented in Fig.~\ref{fig:Fig_jpipi}(right) in the~region \mbox{$3.67<m(\jpp)<3.7\gevcc$}.

The background subtraction is performed with the \sPlot~technique~\cite{Pivk:2004ty} using
the~$\jpsi\pip\pim\proton\Km$~mass as 
the~discriminating variable.
The~signal yield of $\LbPsijppPK$~decays
 is determined using an unbinned extended maximum likelihood fit to the~$\jpp$ invariant mass distribution.
The~$\psitwos$~component is modelled with a~modified Gaussian function with power\nobreakdash-law tails on
both sides,
where the~tail parameters are fixed from simulation. 
The~nonresonant component is taken to be constant.
The~mass resolution parameter obtained from the~fit is $2.29\pm0.17\mevcc$.
The~signal yields are summarized in Table~\ref{tab:sigfit_stats}.

\begin{table}[t]
  \centering
  \caption{\small Signal yields of \Lb~decay channels. Uncertainties are statistical only.}
  \label{tab:sigfit_stats}
  \begin{center}
    \begin{tabular}{lc}	%{l|c}
      Channel	& $N(\Lb)$ \\
      \hline
      $\LbJpsiPK$	& $28\,834\pm204\,$	\\
      $\LbPsimmPK$	& $\phantom{0}665\pm28$	\\
      $\LbPsijppPK$	& $\phantom{0}231\pm17$ \\
      $\LbJpipiPK$	& $\phantom{0}793\pm36$	
    \end{tabular}
  \end{center}
\end{table}

%\section{Branching fractions} % v0.1 % % Vanya's proposal \label{sec:Branching}

The~ratio of branching fractions $R^{\psitwos}$, defined in Eq.~\ref{eq:rdef1}, is measured in two different decay modes,
\begin{equation}   
\label{eq:ratio}
\left.
\begin{aligned}                
 \left. R^{\psitwos}\right|_{\psitwos\to\mumu} 
& = \dfrac{ N_{\psimm} }{ N_{\jpsi} } \times \dfrac{ \varepsilon^{\Lb}_{\jpsi} }{ 
  \varepsilon^{\Lb}_{\psimm} } \times
  \dfrac{ \BR(\jpsimm) }{\BR(\psimm)} \,,     \\ %% \label{eq:ratioBR1} \\
\left. R^{\psitwos}\right|_{\psitwos\to\jpsi\pip\pim} 
& = \dfrac{ N_{\psijpp} }{ N_{\jpsi} } \times \dfrac{ \varepsilon^{\Lb}_{\jpsi} }{ 
  \varepsilon^{\Lb}_{\psijpp} } \times \dfrac{ 1 }{\BR(\psitwos\to\jpp)} \,,    %% \label{eq:ratioBR2}
\end{aligned}
\right.
\end{equation}
\noindent and the ratio $R^{\jpp}$, defined in Eq.~\ref{eq:rdef2}, is measured as
\begin{equation}
 R^{\jpsi\pip\pim} 
= \dfrac{ N_{\jpp} }{ N_{\jpsi} } \times \dfrac{ \varepsilon^{\Lb}_{\jpsi} }{ \varepsilon^{\Lb}_{\jpp} }\,, % \label{eq:ratioBR3}
\end{equation}
where $N_{\mathrm{X}}$ represents the~observed signal yield and $\varepsilon^{\Lb}_{\mathrm{X}}$ denotes the~efficiency for 
the~decay~\mbox{$\Lb\to\mathrm{X}\proton\Km$}.
The~ratio $\dfrac{ \BR(\jpsimm) }{\BR(\psimm)}$ is taken to be equal to the~more precisely measured ratio of dielectron branching fractions,
$\dfrac{ \BR(\jpsi\to \ep\en) }{\BR(\psitwos \to \ep\en)}=7.57\pm0.17$~\cite{PDG2014}. 
For~the~$\psijpp$ branching fraction the world average \mbox{$(34.46\pm0.30)\%$}~\cite{PDG2014} is taken.

The~efficiency is defined as the~product of the geometric acceptance and the detection, reconstruction, 
selection and trigger efficiencies. 
The~efficiencies for hadron identification as functions of kinematic parameters and event multiplicity are determined 
from data using calibration samples of low-background decays: ${\Dstarp\to\Dz\pip}$ followed by 
${\Dz\to\Km\pim}$ for kaons and pions, and \mbox{$\PLambda^0\to\proton\pim$} and 
\mbox{$\Lc\to\proton\Km\pip$} for protons \cite{LHCb-DP-2012-003}.
The~remaining efficiencies are determined using simulation. 

In~the~simulation of $\LbJpsiPK$~decays, the~model established in Ref.~\cite{LHCb-PAPER-2015-029} 
that includes pentaquark contributions
is used, while in the simulation of the~other decay modes the events are generated uniformly in phase space.
The simulation is corrected to reproduce the~transverse momentum and rapidity distributions of
the~$\Lb$~baryons observed in data~\cite{LHCb-PAPER-2015-032} 
and to account for small discrepancies between data and simulation in the reconstruction of 
charged tracks~\cite{LHCb-DP-2013-002}.
The~ratios of efficiencies to those in the~$\LbJpsiPK$ channel are presented in Table~\ref{tab:EffRat}.

\begin{table}[tb]
\centering
\caption{\small Ratios of efficiencies. The uncertainties reflect the limited size of the simulation sample.}
\begin{center}
{ \renewcommand{\arraystretch}{1.5}
\begin{tabular}{lc}
                 & Value \\
  \hline
	%% Vanya's proposal
  $\varepsilon^{\Lb}_{\jpsi}/\varepsilon^{\Lb}_{\psimm}$	& $1.188\pm0.006$ \\
  $\varepsilon^{\Lb}_{\jpsi}/\varepsilon^{\Lb}_{\psijpp}$	& $8.84\pm0.05$	\\
  $\varepsilon^{\Lb}_{\jpsi}/\varepsilon^{\Lb}_{\jpp}$	& $7.59\pm0.04$

\end{tabular}
}
\end{center}
\label{tab:EffRat}
\end{table}

%%\subsection{Systematic uncertainties and results}  
\subsection{Systematic uncertainties}  
                        
Most systematic uncertainties cancel in the measurements of the~ratios of branching fractions,
notably those related to the~reconstruction, identification and trigger efficiencies of the~\mbox{$\jpsi\to\mumu$} and 
\mbox{$\psitwos\to\mumu$}~candidates~\cite{LHCb-PAPER-2012-010},
due to the~similarity of the~muon and dimuon spectra for these modes.
The~remaining systematic uncertainties are summarized in Table~\ref{table:systBR} and discussed below.

\begin{table}[tb]
\caption{\small Systematic uncertainties (in \%) on the ratios of branching fractions  $R^{\psitwos}$ and $R^{\jpsi\pip\pim}$. 
}
\begin{center}
\begin{tabular}{lccc}

%%	{Source}		&{$\psimm$} 	& {$\psijpp$}	& {$\jpp$}    \\
	{Source}		& $\left.R^{\psitwos}\right|_{\psimm}$
                                & $\left.R^{\psitwos}\right|_{\psijpp}$
                                & $R^{\jpsi\pip\pim}$ \\ 
	\hline
	Fit model		& $0.8$		& $3.0$    	& $3.5$  \\
	Cross-feed		& $0.8$ 	& $0.9$ 	& $0.9$	 \\
%%	\Bs and \Bd veto
        %%
        Efficiency calculation:  &  & & \\
	~~\Lb~decay model 	& $0.3$ 	& $0.8$    	& $0.8$	\\
        \multicolumn{2}{l}{~~Reconstruction of additional pions:}  & & \\ 
%        ~~Pion reconstruction:  &  & & \\ 
	~~~~Hadron interaction		& --    	& $2\times2.0$	& $2\times2.0$	\\
	~~~~Track efficiency correction	& --    	& $3.2$		& $2.7$	\\ % 
	~~Hadron identification		& $0.1$ 	& $0.1$		& $0.2$	\\
	~~Trigger			& $1.1$		& $1.1$		& $1.1$	\\
        ~~Selection criteria		& $0.6$ 	& $0.9$ 	& $0.2$	\\
	%%~~simulation statistic	& $0.5$		& $0.6$		& $0.5$	\\
	~~Simulation sample size	& $1.0$		& $1.6$		& $1.7$	\\
	\hline
        Sum in quadrature 	& $2.0$ 	& $6.4$   	& $6.4$	%
\end{tabular}
\end{center}
\label{table:systBR}
\end{table}

% Fit model

Alternative parametrizations for the~signal and background are used to estimate 
the~systematic uncertainties related to the~fit model. 
A~modified Novosibirsk function~\cite{Lees:2011gw}, 
an~Apolonios function~\cite{Apolonios}, 
an~asymmetric variant of the~Apolonios function 
and the~Student's t\nobreakdash-distribution 
are used 
for the~\Lb~signal shape,
and an~exponential function multiplied by a~second\nobreakdash-order 
polynomial is used 
for the background. 
The~ratio of event yields is remeasured with the~cross\nobreakdash-check models, and the~maximum deviation with
respect to the~nominal value is assigned as a~systematic uncertainty.

% Mass veto

The~uncertainty associated with the~\Bs and \Bd~cross\nobreakdash-feed is estimated by varying the~widths of the~rejected regions and recomputing 
the~signal yields, taking into account the~changes in efficiencies.
As~an additional cross\nobreakdash-check,
a~veto is applied also on possible contributions from 
\mbox{$\Lbbar\to\jpsi\antiproton\Kp$}, \mbox{$\Lbbar\to\psitwos\antiproton\Kp$} and \mbox{$\Lbbar\to\jpsi\pip\pim\antiproton\Kp$}~decays
where the~positive kaon is misidentified as a~proton and the~antiproton is misidentified as a~negative kaon.  
The~maximum of the~observed differences from the~nominal values
is assigned as the~corresponding systematic uncertainty.

The remaining systematic uncertainties are associated with the efficiency determination.
%%
% Resonance structure
%%
The~systematic uncertainty related to the~decay model for
\mbox{$\Lb\to\psitwos\proton\Km$} and \mbox{$\Lb\to\jpsi\pip\pim\proton\Km$}~decays
is estimated using the~simulated samples, 
corrected to reproduce the~invariant mass of the~$\proton\Km$ and $\psitwos\proton$ or $\jpp\proton$~systems observed in data.
The~largest change in efficiency is 
taken as the~corresponding systematic uncertainty.

% Reweighting on Lb kinematics

% Two additional tracks

The~decay modes \mbox{\LbJpipiPK}~and \mbox{$\Lb\to\psitwos[\to\jpp]\proton\Km$}~have two 
additional pions to reconstruct compared to the~reference mode~\LbJpsiPK.
The~uncertainty associated with the reconstruction of these additional
low-\pt~tracks has two independent contributions.
First, the uncertainties in the amount and distribution of material in~the~detector result in an~uncertainty
of 2.0\%~per additional final\nobreakdash-state pion due to the modelling of hadron interactions~\cite{LHCb-DP-2013-002}.
Second, the~small difference in the~track finding efficiency between data and simulation
is corrected using a data\nobreakdash-driven technique~\cite{LHCb-DP-2013-002}.
The~uncertainties in the~correction factors
are propagated to the~efficiency ratios
by means of pseudoexperiments.
This~results in a~systematic uncertainty of $3.2\%$ for the 
ratio  $\left.R^{\psitwos}\right|_{\psijpp}$
and $2.7\%$ for the~ratio  $R^{\jpsi\pip\pim}$.

% Hadron identification

The~systematic uncertainties related to the hadron % proton, kaon and pion 
identification efficiency,
\mbox{0.1\,(0.2)\%} for \mbox{$R^{\psitwos}\,(R^{\jpsi\pip\pim})$} ratios, 
reflect
the~limited sizes of the~calibration samples,
and are propagated to the~ratios $R^{\psitwos}$
and $R^{\jpsi\pip\pim}$ by means of pseudoexperiments.

% Trigger

The trigger efficiency for events with $\jpsimm$ and $\psimm$ produced in beauty hadron decays is studied in data. A systematic uncertainty of 1.1\% 
is assigned based on a comparison
between data and simulation of the ratio of trigger efficiencies for high-yield samples
of $\Bp\to\jpsi\Kp$ and $\Bp\to\psitwos\Kp$ decays~\cite{LHCb-PAPER-2012-010}.

% Data-simulation agreement

Another source of uncertainty is the~potential disagreement between data and simulation 
in the~estimation of efficiencies, due to effects not considered above. 
This~is studied by varying the~selection criteria 
in ranges that lead to as much as $\pm20\%$ change in the~measured signal yields. 
The~stability is tested by comparing the efficiency-corrected yields within these variations. 
The largest deviations range between 0.2\% and 0.9\% and 
are taken as systematic uncertainties.

% MC statistic

Finally, a systematic uncertainty due to the limited size of the simulation sample is assigned.
With all the systematic uncertainties added in quadrature,
the total is
$2.0\%$ for the ratio $\left.R^{\psitwos}\right|_{\psimm}$,
$6.4\%$ for the ratio $\left.R^{\psitwos}\right|_{\psijpp}$ 
and $6.4\%$ for the~ratio~$R^{\jpp}$.

%%%%%%%%%%% The results %%%%%%%%%%%%%%%%

%\subsection{Results}
\subsection{Results}

Using Eq.~\ref{eq:ratio} and the ratios of yields and efficiencies determined above,
the~ratio~$R^{\psitwos}$
is measured for each \psitwos decay mode separately:
%%\begin{subequations}
\begin{equation}\label{eq:result1}
\left. 
  \begin{aligned}
   \left.R^{\psitwos}\right|_{\psitwos\to\mumu}
    &= (20.74\pm 0.88 \pm 0.41 \pm 0.47 )\times10^{-2}\,,     \\ %%   \label{eq:result1a}\\
    \left.R^{\psitwos}\right|_{\psitwos\to\jpsi\pip\pim}    
    &= (20.55\pm 1.52 \pm 1.32 \pm 0.18 )\times10^{-2}\,,     \\ %%  \label{eq:result1b}
  \end{aligned}
  \right. 
\end{equation}
where the~first uncertainty is statistical, the~second is systematic and the~third
is related to the uncertainties on the dielectron \jpsi and \psitwos branching fractions and
the~$\psijpp$ branching fraction.
The~average of the~ratios in Eq.~\ref{eq:result1} is %% calculated  to be 
\begin{equation}
    R^{\psitwos}
= (20.70\pm 0.76 \pm 0.46 \pm 0.37)\times10^{-2}\,.
\end{equation}
In this average the~systematic uncertainties related to
the~normalization channel, $\LbJpsiPK$,
and the~trigger efficiency are considered
to be 100\% correlated while other
systematic uncertainties  are treated as uncorrelated. 

The ratio of the branching fractions of $\LbJpipiPK$ and $\LbJpsiPK$ is found to be 
\begin{equation}
  R^{\jpsi\pip\pim} = (20.86\pm 0.96 \pm 1.34)\times10^{-2}\,, 
  \label{eq:result2}
\end{equation}
where contributions via intermediate resonances are included.

The~absolute branching fractions
\LbPsiPK~and \LbJpipiPK
are derived
using the branching fraction
\mbox{$\BR(\LbJpsiPK)=(3.04\pm0.04\pm0.06\pm0.33\,^{+0.43}_{-0.27})\times 10^{-4}$},
measured
in Ref.~\cite{LHCb-PAPER-2015-032},
where the~third uncertainty is due to the~uncertainty
on the~branching fraction of the decay $\Bz \to \jpsi \overline{\mathrm{K}}^{*}(892)^0$
and the fourth is due to the knowledge of the~ratio of fragmentation fractions
$f_{\Lb}/f_{\dquark}$. They are found to be 
\begin{equation}
\left.\begin{aligned}
\BR(\LbPsiPK)   &= (6.29\pm 0.23 \pm 0.14\,^{+1.14}_{-0.90} )\times10^{-5}\,, \\ 
\BR(\LbJpipiPK) &= (6.34\pm 0.29 \pm 0.41\,^{+1.15}_{-0.91} )\times10^{-5}\,, 
\end{aligned}
\right.
\end{equation}
where the~third uncertainty comes from
the~uncertainties in
the~branching fractions
of $\LbJpsiPK$, $\psijpp$,
$\psitwos\to\ep\en$ and
$\jpsi\to\ep\en$~decays.

%%%%%%%%%%%%%%%%%%%%%%%%%%%%%%%%%%%%%%%%%%%

From the~two separate measurements of
the~ratio  $R^{\psitwos}$ via different decay modes of the~\psitwos~meson\,(Eq.~\ref{eq:result1}),
the ratio of the~$\psimm$ and $\psijpp$ branching fractions is calculated as 
\begin{align}
  \dfrac{\BR(\psimm)}{\BR(\psijpp)} 
  &= \dfrac{N_{\psimm}}{N_{\psijpp}} \times \dfrac{ \varepsilon^{\Lb}_{\psijpp} }{
    \varepsilon^{\Lb}_{\psimm} } \times \BR(\jpsimm)   \nonumber \\
  &= (2.30\pm0.20 \pm0.12 \pm0.01)\times10^{-2} \,,
\end{align}
where  the~third uncertainty is related to
the~uncertainty of the~known branching
fraction \mbox{$\BR(\jpsimm)=(5.961\pm0.033)\%$}~\cite{PDG2014}. 
This~result is in agreement with the~world average of \mbox{$(2.29\pm0.25)\times10^{-2}$}~\cite{PDG2014}
based on results of the E672/E706~\cite{BrPsi_E672} and 
\babar~\cite{BrPsi_babar} collaborations,
and has similar precision.

\section{Measurement of \Lb baryon mass}
\label{sec:Mass}

\begin{figure}[t]
  \setlength{\unitlength}{1mm}
  \centering
  \begin{picture}(150,67)
    \put(0,2){\includegraphics*[width=75mm,height=65mm]{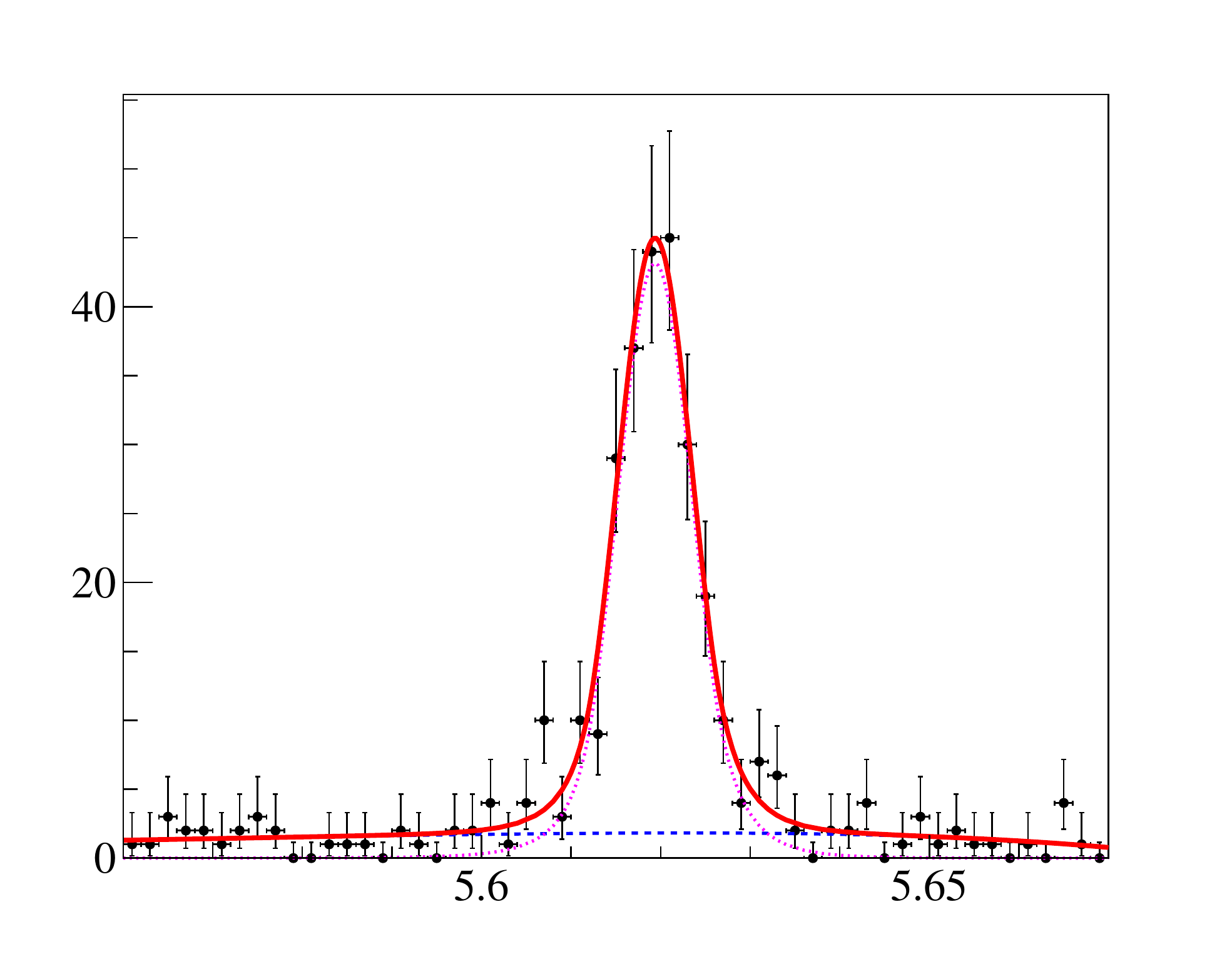}}
    \put( -5,20 )  { \begin{sideways}Candidates/(2$\mevcc$)\end{sideways}  }
    %%\put( 15,0  )  { $m(\jpp\proton\Km)$ }
    \put( 20,0  )  { $m(\psitwos \proton\Km)$}
    \put(80,2){\includegraphics*[width=75mm,height=65mm]{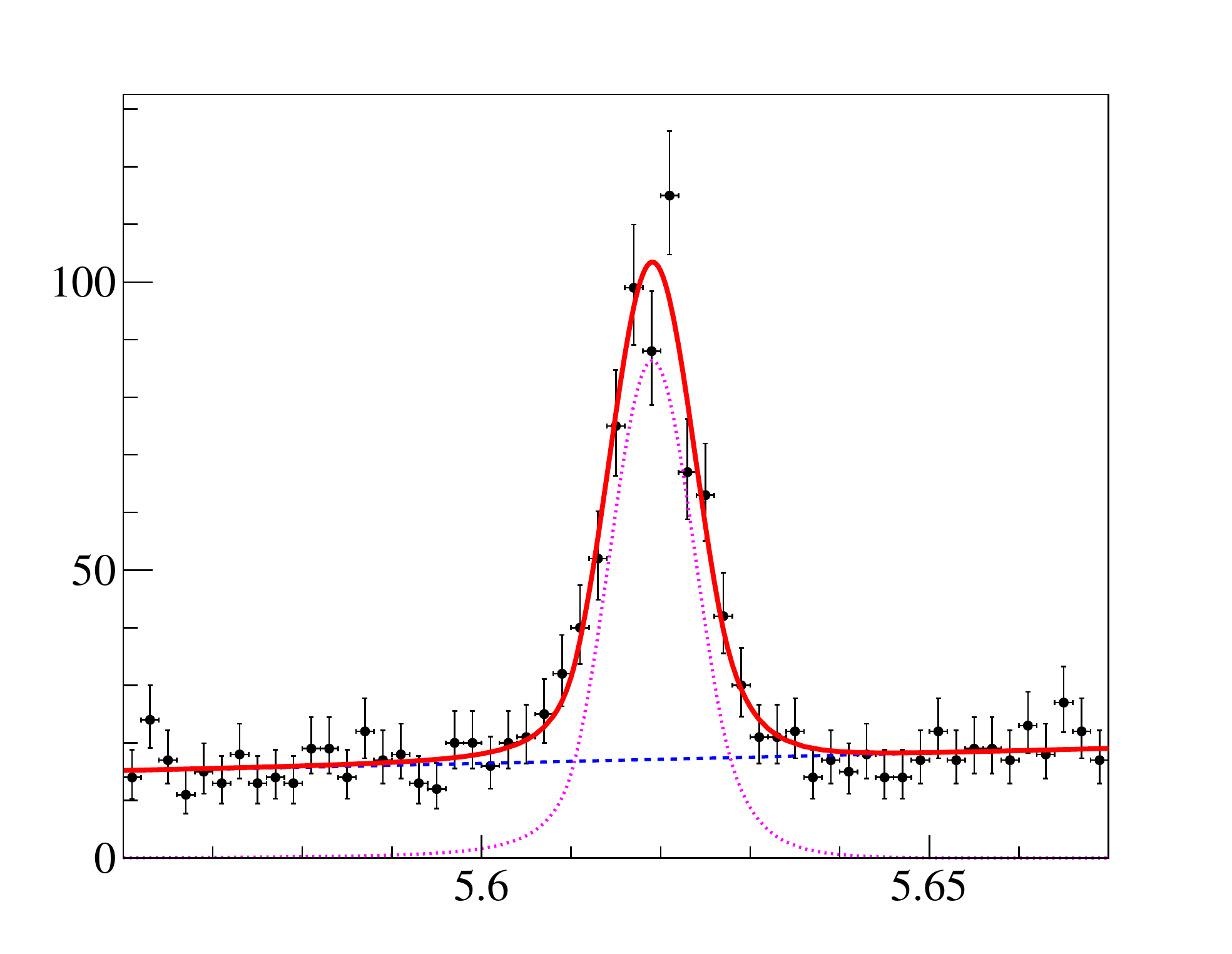}}
    \put( 75,20  )  { \begin{sideways}Candidates/(2$\mevcc$)\end{sideways}  }
    \put( 97,0  )  { $m(\jpsi\pip\pim\proton\Km)$ }
    \put( 51,0  )  { $\left[\gevcc\right]$ }
    \put( 131,0 )  { $\left[\gevcc\right]$ }
    \put( 10,53  ){ \small LHCb}
    \put( 90,53  ){ \small LHCb}
  \end{picture}
  \caption { \small
    (left)~Mass distribution of selected
      $\LbPsijppPK$ candidates  with an~additional constraint for the~\psitwos~mass~\cite{PDG2014}.
    (right)~Mass distribution of selected
      $\LbJpipiPK$~candidates with  a~requirement of the~$\jpp$ combination mass to be outside the~range \mbox{$3670<m(\jpp)<3700\mevcc$}.
    The~total fit 
    function\,(solid red),
    the~$\Lb$ signal contribution\,(dotted magenta)
    and the~combinatorial background\,(dashed blue) are shown.
    }
  \label{fig:Fig_mass2}
\end{figure}

The low energy release in $\LbPsiPK$ and $\LbJpipiPK$~decays allows 
the~\Lb mass 
to be determined
with a~small systematic uncertainty.
The~mass is measured using four decay channels:
$\LbPsimmPK$,
$\LbPsijppPK$,
$\LbJpipiPK$ and
$\LbJpsiPK$.
The~mass distributions for the $\LbPsimmPK$ and $\LbJpsiPK$ channels are shown in Fig.~\ref{fig:Fig_psimm}.
In~the~$\LbPsijppPK$ channel, the~$\jpp$ system is constrained to the~nominal \psitwos
mass~\cite{PDG2014} to improve the precision. 
In~the~$\LbJpipiPK$ channel, to avoid overlap with the~$\LbPsijppPK$ channel
the~\psitwos region is vetoed,
\ie the~mass of the~$\jpp$ combination is required to be outside the~range \mbox{$3670<m(\jpp)<3700\mevcc$}.
The~mass distributions for these two samples, along with
    the result  of an~unbinned extended maximum likelihood fit
    using the model described in Sect.~\ref{sec:signalpeaks},
    are shown in Fig.~\ref{fig:Fig_mass2}.

The systematic uncertainties on the measurement of the \Lb baryon mass for all four 
channels are listed in Table~\ref{table:Mass_syst}. 
The~precision of the absolute momentum scale calibration of $0.03\%$ is the~dominant source of 
uncertainty~\cite{LHCb-PAPER-2012-048,LHCb-PAPER-2013-011}.
This~uncertainty is proportional to the~energy release in the~decay and is minimal for
    the~processes with a~\psitwos in the~final state.
A~further uncertainty is related to the~energy loss in the~material of the~tracking system~\cite{LHCb-PAPER-2011-035}, 
which is known with an~accuracy of 10\%~\cite{LHCb-PAPER-2010-001}.
This~effect is estimated  by varying %% the~magnitude of 
the~energy loss correction in the~reconstruction by 10\% and taking the~observed mass shift as an~uncertainty.
The~uncertainty due to the~fit model is estimated using the~same 
set of cross\nobreakdash-check models for the~signal and background parameterization as considered in Sect.~\ref{sec:signals}, with
the~maximum deviation in the~mass assigned as a~systematic uncertainty.
The~uncertainties on the~masses of the~\jpsi and \psitwos~mesons~\cite{PDG2014} are small
and are therefore neglected.

As a~cross-check, the~data sample is divided into four parts,
for data collected at $\sqrt{s}=7$ and $8\tev$ and with different magnet polarities.
The~measured masses are consistent among these subsamples, and therefore no systematic uncertainty is assigned. 
To~check the effect of the~selection criteria\,(see Sect.~\ref{sec:Selection}), 
the high\nobreakdash-yield~\mbox{$\LbJpsiPK$}~decay channel is used. 
No~sizeable dependence of the mass on the selection criteria 
is observed and no additional uncertainty is assigned.

\begin{table}[t]
  \caption{\small Systematic uncertainties (in \mevcc) on the \Lb mass using
    the~decay modes~$\LbJpsiPK$, $\LbPsimmPK$, 
    $\LbPsijppPK$ and $\LbJpipiPK$ with the $\jpp$~mass outside the~\psitwos~region.
  }
  \begin{center}
    \begin{tabular}{lcccc}
      & $\jpsi$	& $\psimm$	& $\psijpp$  & $\jpp,\cancel{\psitwos}$ \\
      \hline
      Momentum scale			& $0.34$    & $0.19$    & $0.15$       & $0.26$	\\
      Energy loss correction  	& $0.03$    & $0.02$    & $0.06$       & $0.07$	\\
      Fit model   			& $0.04$    & $0.03$    & $0.08$       & $0.05$ \\
      \hline
      Sum in quadrature 		& $0.34$    & $0.19$    & $0.18$       & $0.27$
    \end{tabular}
  \end{center}
  \label{table:Mass_syst}
\end{table}

The results from the four decay channels are
presented in Table~\ref{table:Mass_results}. 
To~combine them, correlations must be taken into account.
The~statistical uncertainties and those related 
to the~fit procedure are treated as uncorrelated while those due to the momentum scale and energy loss
correction are considered to be fully correlated. 
The~combined value of the~\Lb~mass is
\begin{equation}
   M(\Lb) = 5619.65 \pm 0.17 \pm 0.17 \mevcc \,,
\end{equation}
where the first uncertainty is statistical and the second systematic.
The~$\chisq{\rm /ndf}$ calculated for the individual measurements with respect to the combined value is 3.0/3.
This is the most precise measurement of any $\bquark$-hadron mass reported to date.

\begin{table}[t]
  \caption{\small Measured \Lb mass in different decay channels.
    The first uncertainty is statistical and the second is systematic.
  }
  \begin{center}
  { \renewcommand{\arraystretch}{1.2}
    \begin{tabular}{lc}
      Channel   & $M(\Lb)\left[\mevcc\right]$   \\
      \hline
      
      $\LbJpsiPK$  			& $5619.62\pm 0.04\pm 0.34$ \\
      $\LbPsimmPK$ 			& $5619.84\pm 0.18\pm 0.19$ \\
      $\LbPsijppPK$   		& $5619.38\pm 0.33\pm 0.18$ \\
      $\LbJpipiPK$ excluding $\psitwos$	& $5619.08\pm 0.30\pm 0.27$ \\
    \end{tabular}
  }
  \end{center}
  \label{table:Mass_results}
\end{table}

Previous direct measurements of the~$\Lb$ mass by \lhcb were made using 
the~decay~\mbox{${\Lb\to\jpsi\PLambda^0}$}~\cite{LHCb-PAPER-2011-035,LHCb-PAPER-2012-048} and 
are statistically independent of the~results of this study. 
The~combination obtained here is consistent with, and more precise than, the results of these earlier studies.
The \lhcb results are combined, taking 
the statistical uncertainties and those related to the fit procedure to be uncorrelated 
and those due to the~energy loss correction to be fully correlated.
The~uncertainty due to the~momentum scale in Ref.~\cite{LHCb-PAPER-2012-048}
is also taken to be fully correlated,
whereas in Ref.~\cite{LHCb-PAPER-2011-035}
a~different alignment and calibration procedure was used and
so the~corresponding uncertainty is considered to be uncorrelated
with the~other measurements.
The~result of the combination is dominated by the~measurements of this analysis and is
\begin{equation}
M(\Lb) = 5619.65 \pm 0.16 \pm 0.14 \mevcc\,,  \label{eq:mlb}
\end{equation}
where the uncertainties are statistical and systematic.
The~$\chisq{\rm /ndf}$ calculated for the individual measurements with respect to the combined value is 3.4/5.
The~measured  mass is in agreement with,
but more precise than, the~results of the \atlas~\cite{LbMassAtlas} and \cdf~\cite{LbMassCDF}~collaborations.

From the value of the $\Lb$ mass in Eq.~\ref{eq:mlb} and a~precise measurement of
the~mass difference between 
the $\Lb$ and $\Bd$ hadrons reported in~Ref.\cite{LHCb-PAPER-2014-002}, 
the~mass of the \Bd~meson is calculated to be 
\begin{equation}
  M(\Bd) = 5279.93 \pm 0.39 \mevcc\,,
\end{equation}
where the correlation of 41\% between the LHCb measurements of the $\Lb$ mass 
and the~\mbox{$\Lb$--$\Bd$}~mass splitting has been taken into account. 
This is in agreement with the~current world average of $5279.61\pm0.16\mevcc$~\cite{PDG2014}.

\section{Results and summary}
\label{sec:Result}

The~$\LbPsiPK$ and $\LbJpipiPK$ decay modes are observed
using a~sample of $\proton\proton$ collisions at centre-of-mass energies of 7 and 8\tev,
corresponding to an integrated luminosity of 3\invfb.
With results from  the channels $\psimm$ and $\psijpp$~combined, the~ratio
of branching fractions is  measured:
\begin{equation*}
 R^{\psitwos} = \dfrac{\BR(\LbPsiPK)}{\BR(\LbJpsiPK)} = (20.70\pm 0.76\pm 0.46\pm 0.37)\times10^{-2}\,,
\end{equation*}
where the first uncertainty is statistical, the second is systematic and the third is
related to the~uncertainties of the~known dielectron $\jpsi$ and 
$\psitwos$ branching fractions and of the~branching fraction of
the~$\psijpp$ decay.
The~ratio of branching fractions for $\LbJpipiPK$ and $\LbJpsiPK$ is
\begin{equation*}
  R^{\jpsi\pip\pim} = 
  \dfrac{\BR(\LbJpipiPK)}{\BR(\LbJpsiPK)}  = (20.86\pm 0.96\pm 1.34)\times10^{-2}\,, 
\end{equation*}
where the first uncertainty is statistical, the second is systematic and
contributions via intermediate resonances are included. 

From measurements of the~ratio $R^{\psitwos}$ via two different decay modes of the \psitwos~meson
it is determined that
\begin{equation*}
  \dfrac{\BR(\psimm)}{\BR(\psijpp)} = (2.30\pm0.20\pm0.12\pm0.01)\times10^{-2} \,,
\end{equation*}
\noindent where the first uncertainty is statistical, the second is systematic and 
the~third is related to the uncertainty on $\BR(\jpsi\to\mumu)$.
This~is the most precise direct measurement of this ratio to date.

%The 

Using $\LbPsiPK$,
$\LbJpipiPK$ and 
$\LbJpsiPK$~decays, the~mass of the~\Lb~baryon is measured to be
\begin{equation*}
  M(\Lb) = 5619.65 \pm 0.17 \pm 0.17 \mevcc\,,
\end{equation*}
where the first uncertainty is statistical and the second is systematic. 
Combining this result with previous \lhcb measurements that used the channel ${\Lb\to\jpsi\Lambda^0}$~\cite{LHCb-PAPER-2011-035,LHCb-PAPER-2012-048} 
gives
\begin{equation}
  M(\Lb) = 5619.65 \pm 0.16 \pm 0.14 \mevcc\,,
\end{equation}
where the first uncertainty is statistical and the second is systematic.
This~is the most precise determination of the mass of any $\bquark$ hadron to date.

% Do not include this in analysis note and conference reports
\section*{Acknowledgements}

\noindent We express our gratitude to our colleagues in the~CERN
accelerator departments for the~excellent performance of the~LHC.
We~thank the technical and administrative staff at the~LHCb
institutes.
We~acknowledge support from CERN and from the~national
agencies:
CAPES, CNPq, FAPERJ and FINEP\,(Brazil);
NSFC\,(China);
CNRS/IN2P3\,(France);
BMBF, DFG and MPG\,(Germany);
INFN\,(Italy); 
FOM and NWO\,(The~Netherlands);
MNiSW and NCN\,(Poland);
MEN/IFA\,(Romania); 
MinES and FANO\,(Russia);
MinECo\,(Spain);
SNSF and SER\,(Switzerland); 
NASU\,(Ukraine);
STFC\,(United Kingdom);
NSF\,(USA).
We~acknowledge the~computing resources that are provided by CERN,
IN2P3\,(France),
KIT and DESY\,(Germany),
INFN\,(Italy),
SURF\,(The~Netherlands),
PIC\,(Spain),
GridPP\,(United Kingdom),
RRCKI and Yandex LLC\,(Russia),
CSCS\,(Switzerland),
IFIN\nobreakdash-HH\,(Romania),
CBPF\,(Brazil),
PL\nobreakdash-GRID\,(Poland) and OSC\,(USA).
We~are indebted to the~communities behind the~multiple open 
source software packages on which we depend.
Individual groups or members have received support from
AvH Foundation\,(Germany),
EPLANET, Marie Sk\l{}odowska\nobreakdash-Curie Actions and ERC\,(European Union), 
Conseil G\'{e}n\'{e}ral de Haute\nobreakdash-Savoie, Labex ENIGMASS and OCEVU, 
R\'{e}gion Auvergne\,(France),
RFBR and Yandex LLC\,(Russia), GVA, XuntaGal and GENCAT\,(Spain),
Herchel Smith Fund, The~Royal Society, Royal Commission for
the~Exhibition of 1851 and the~Leverhulme Trust\,(United Kingdom).

\addcontentsline{toc}{section}{References}
\setboolean{inbibliography}{true}
\bibliographystyle{LHCb}
\bibliography{main,local,LHCb-PAPER,LHCb-CONF,LHCb-DP,LHCb-TDR}

\ifx\mcitethebibliography\mciteundefinedmacro
\PackageError{LHCb.bst}{mciteplus.sty has not been loaded}
{This bibstyle requires the use of the mciteplus package.}\fi
\providecommand{\href}[2]{#2}
\begin{mcitethebibliography}{10}
\mciteSetBstSublistMode{n}
\mciteSetBstMaxWidthForm{subitem}{\alph{mcitesubitemcount})}
\mciteSetBstSublistLabelBeginEnd{\mcitemaxwidthsubitemform\space}
{\relax}{\relax}

\bibitem{Neubert:1993mb}
M.~Neubert, \ifthenelse{\boolean{articletitles}}{\emph{{Heavy quark symmetry}},
  }{}\href{http://dx.doi.org/10.1016/0370-1573(94)90091-4}{Phys.\ Rept.\
  \textbf{245} (1994) 259},
  \href{http://arxiv.org/abs/hep-ph/9306320}{{\normalfont\ttfamily
  arXiv:hep-ph/9306320}}\relax
\mciteBstWouldAddEndPuncttrue
\mciteSetBstMidEndSepPunct{\mcitedefaultmidpunct}
{\mcitedefaultendpunct}{\mcitedefaultseppunct}\relax
\EndOfBibitem
\bibitem{Manohar:2000dt}
A.~V. Manohar and M.~B. Wise, \ifthenelse{\boolean{articletitles}}{\emph{{Heavy
  quark physics}}, }{}Camb.\ Monogr.\ Part.\ Phys.\ Nucl.\ Phys.\ Cosmol.\
  \textbf{10} (2000) 1\relax
\mciteBstWouldAddEndPuncttrue
\mciteSetBstMidEndSepPunct{\mcitedefaultmidpunct}
{\mcitedefaultendpunct}{\mcitedefaultseppunct}\relax
\EndOfBibitem
\bibitem{Abulencia:2006df}
CDF collaboration, A.~Abulencia {\em et~al.},
  \ifthenelse{\boolean{articletitles}}{\emph{{Measurement of
  $\sigma(\Lb)/\sigma(\bar{\B}^0) \times \BR(\Lb \to \Lc \pim)/\BR(\bar{\B}^0
  \to \D^+ \pim)$ in $\proton\antiproton$~collisions at $\sqs=1.96\tev$}},
  }{}\href{http://dx.doi.org/10.1103/PhysRevLett.98.122002}{Phys.\ Rev.\ Lett.\
   \textbf{98} (2007) 122002},
  \href{http://arxiv.org/abs/hep-ex/0601003}{{\normalfont\ttfamily
  arXiv:hep-ex/0601003}}\relax
\mciteBstWouldAddEndPuncttrue
\mciteSetBstMidEndSepPunct{\mcitedefaultmidpunct}
{\mcitedefaultendpunct}{\mcitedefaultseppunct}\relax
\EndOfBibitem
\bibitem{Abazov:2011wt}
D0 collaboration, V.~M. Abazov {\em et~al.},
  \ifthenelse{\boolean{articletitles}}{\emph{{Measurement of the production
  fraction times branching fraction $f(\bquark\to\Lb)\cdot \mathcal{B}(\Lb\to
  \jpsi \Lambda)$}},
  }{}\href{http://dx.doi.org/10.1103/PhysRevD.84.031102}{Phys.\ Rev.\
  \textbf{D84} (2011) 031102},
  \href{http://arxiv.org/abs/1105.0690}{{\normalfont\ttfamily
  arXiv:1105.0690}}\relax
\mciteBstWouldAddEndPuncttrue
\mciteSetBstMidEndSepPunct{\mcitedefaultmidpunct}
{\mcitedefaultendpunct}{\mcitedefaultseppunct}\relax
\EndOfBibitem
\bibitem{LbMassCDF}
\cdf collaboration, T.~Aaltonen {\em et~al.},
  \ifthenelse{\boolean{articletitles}}{\emph{{Mass and lifetime measurements of
  bottom and charm baryons in $\proton\antiproton$ collisions at
  $\sqrt{s}=1.96\tev$}},
  }{}\href{http://dx.doi.org/10.1103/PhysRevD.89.072014}{Phys.\ Rev.\
  \textbf{D89} (2014) 072014},
  \href{http://arxiv.org/abs/1403.8126}{{\normalfont\ttfamily
  arXiv:1403.8126}}\relax
\mciteBstWouldAddEndPuncttrue
\mciteSetBstMidEndSepPunct{\mcitedefaultmidpunct}
{\mcitedefaultendpunct}{\mcitedefaultseppunct}\relax
\EndOfBibitem
\bibitem{LHCb-PAPER-2013-056}
LHCb collaboration, R.~Aaij {\em et~al.},
  \ifthenelse{\boolean{articletitles}}{\emph{{Study of beauty baryon decays to
  $\D^0 \proton \mathrm{h}^-$ and $\Lc \mathrm{h}^-$ final states}},
  }{}\href{http://dx.doi.org/10.1103/PhysRevD.89.032001}{Phys.\ Rev.\
  \textbf{D89} (2014) 032001},
  \href{http://arxiv.org/abs/1311.4823}{{\normalfont\ttfamily
  arXiv:1311.4823}}\relax
\mciteBstWouldAddEndPuncttrue
\mciteSetBstMidEndSepPunct{\mcitedefaultmidpunct}
{\mcitedefaultendpunct}{\mcitedefaultseppunct}\relax
\EndOfBibitem
\bibitem{LHCb-PAPER-2014-002}
LHCb collaboration, R.~Aaij {\em et~al.},
  \ifthenelse{\boolean{articletitles}}{\emph{{Study of beauty hadron decays
  into pairs of charm hadrons}},
  }{}\href{http://dx.doi.org/10.1103/PhysRevLett.112.202001}{Phys.\ Rev.\
  Lett.\  \textbf{112} (2014) 202001},
  \href{http://arxiv.org/abs/1403.3606}{{\normalfont\ttfamily
  arXiv:1403.3606}}\relax
\mciteBstWouldAddEndPuncttrue
\mciteSetBstMidEndSepPunct{\mcitedefaultmidpunct}
{\mcitedefaultendpunct}{\mcitedefaultseppunct}\relax
\EndOfBibitem
\bibitem{LHCb-PAPER-2014-020}
LHCb collaboration, R.~Aaij {\em et~al.},
  \ifthenelse{\boolean{articletitles}}{\emph{{Observation of the $\Lb \to \jpsi
  \proton \pim$ decay}},
  }{}\href{http://dx.doi.org/10.1007/JHEP07(2014)103}{JHEP \textbf{07} (2014)
  103}, \href{http://arxiv.org/abs/1406.0755}{{\normalfont\ttfamily
  arXiv:1406.0755}}\relax
\mciteBstWouldAddEndPuncttrue
\mciteSetBstMidEndSepPunct{\mcitedefaultmidpunct}
{\mcitedefaultendpunct}{\mcitedefaultseppunct}\relax
\EndOfBibitem
\bibitem{LHCb-PAPER-2015-032}
LHCb collaboration, R.~Aaij {\em et~al.},
  \ifthenelse{\boolean{articletitles}}{\emph{{Study of the productions of $\Lb$
  and $\bar{B}^0$ hadrons in $\proton\proton$ collisions and first measurement
  of the $\Lb \to \jpsi \proton \Km$ branching fraction}},
  }{}\href{http://dx.doi.org/10.1088/1674-1137/40/1/011001}{Chin.\ Phys.\ C
  \textbf{40} (2015) 011001},
  \href{http://arxiv.org/abs/1509.00292}{{\normalfont\ttfamily
  arXiv:1509.00292}}\relax
\mciteBstWouldAddEndPuncttrue
\mciteSetBstMidEndSepPunct{\mcitedefaultmidpunct}
{\mcitedefaultendpunct}{\mcitedefaultseppunct}\relax
\EndOfBibitem
\bibitem{LHCb-PAPER-2014-003}
LHCb collaboration, R.~Aaij {\em et~al.},
  \ifthenelse{\boolean{articletitles}}{\emph{{Precision measurement of the
  ratio of the $\Lb$ to $\overline{\B}^0$ lifetimes}},
  }{}\href{http://dx.doi.org/10.1016/j.physletb.2014.05.021}{Phys.\ Lett.\
  \textbf{B734} (2014) 122},
  \href{http://arxiv.org/abs/1402.6242}{{\normalfont\ttfamily
  arXiv:1402.6242}}\relax
\mciteBstWouldAddEndPuncttrue
\mciteSetBstMidEndSepPunct{\mcitedefaultmidpunct}
{\mcitedefaultendpunct}{\mcitedefaultseppunct}\relax
\EndOfBibitem
\bibitem{LHCb-PAPER-2014-048}
LHCb collaboration, R.~Aaij {\em et~al.},
  \ifthenelse{\boolean{articletitles}}{\emph{{Precision measurement of the mass
  and lifetime of the $\Xi_{\bquark}^-$ baryon}},
  }{}\href{http://dx.doi.org/10.1103/PhysRevLett.113.242002}{Phys.\ Rev.\
  Lett.\  \textbf{113} (2014) 242002},
  \href{http://arxiv.org/abs/1409.8568}{{\normalfont\ttfamily
  arXiv:1409.8568}}\relax
\mciteBstWouldAddEndPuncttrue
\mciteSetBstMidEndSepPunct{\mcitedefaultmidpunct}
{\mcitedefaultendpunct}{\mcitedefaultseppunct}\relax
\EndOfBibitem
\bibitem{LHCb-PAPER-2015-029}
LHCb collaboration, R.~Aaij {\em et~al.},
  \ifthenelse{\boolean{articletitles}}{\emph{{Observation of $\jpsi \proton$
  resonances consistent with pentaquark states in $\Lb \to \jpsi \proton \Km$
  decays}}, }{}\href{http://dx.doi.org/10.1103/PhysRevLett.115.072001}{Phys.\
  Rev.\ Lett.\  \textbf{115} (2015) 072001},
  \href{http://arxiv.org/abs/1507.03414}{{\normalfont\ttfamily
  arXiv:1507.03414}}\relax
\mciteBstWouldAddEndPuncttrue
\mciteSetBstMidEndSepPunct{\mcitedefaultmidpunct}
{\mcitedefaultendpunct}{\mcitedefaultseppunct}\relax
\EndOfBibitem
\bibitem{LHCb-PAPER-2012-010}
LHCb collaboration, R.~Aaij {\em et~al.},
  \ifthenelse{\boolean{articletitles}}{\emph{{Measurement of relative branching
  fractions of $\B$ decays to $\psitwos$ and $\jpsi$ mesons}},
  }{}\href{http://dx.doi.org/10.1140/epjc/s10052-012-2118-7}{Eur.\ Phys.\ J.\
  \textbf{C72} (2012) 2118},
  \href{http://arxiv.org/abs/1205.0918}{{\normalfont\ttfamily
  arXiv:1205.0918}}\relax
\mciteBstWouldAddEndPuncttrue
\mciteSetBstMidEndSepPunct{\mcitedefaultmidpunct}
{\mcitedefaultendpunct}{\mcitedefaultseppunct}\relax
\EndOfBibitem
\bibitem{LHCb-PAPER-2012-053}
LHCb collaboration, R.~Aaij {\em et~al.},
  \ifthenelse{\boolean{articletitles}}{\emph{{Observations of $\Bs \to \psitwos
  \Peta$ and \mbox{$\B^0_{(\squark)} \to \psitwos \pip\pim$} decays}},
  }{}\href{http://dx.doi.org/10.1016/j.nuclphysb.2013.03.004}{Nucl.\ Phys.\
  \textbf{B871} (2013) 403 },
  \href{http://arxiv.org/abs/1302.6354}{{\normalfont\ttfamily
  arXiv:1302.6354}}\relax
\mciteBstWouldAddEndPuncttrue
\mciteSetBstMidEndSepPunct{\mcitedefaultmidpunct}
{\mcitedefaultendpunct}{\mcitedefaultseppunct}\relax
\EndOfBibitem
\bibitem{LHCb-PAPER-2013-024}
LHCb collaboration, R.~Aaij {\em et~al.},
  \ifthenelse{\boolean{articletitles}}{\emph{{Observation of $\Bs
  \to\Pchi_{\cquark1}\phi$ decay and study of
  $\Bz\to\Pchi_{\cquark1,2}\mathrm{K}^{*0}$ decays}},
  }{}\href{http://dx.doi.org/10.1016/j.nuclphysb.2013.06.005}{Nucl.\ Phys.\
  \textbf{B874} (2013) 663},
  \href{http://arxiv.org/abs/1305.6511}{{\normalfont\ttfamily
  arXiv:1305.6511}}\relax
\mciteBstWouldAddEndPuncttrue
\mciteSetBstMidEndSepPunct{\mcitedefaultmidpunct}
{\mcitedefaultendpunct}{\mcitedefaultseppunct}\relax
\EndOfBibitem
\bibitem{LHCb-PAPER-2014-050}
LHCb collaboration, R.~Aaij {\em et~al.},
  \ifthenelse{\boolean{articletitles}}{\emph{{Measurement of $\Bc$ production
  at $\sqs=8\tev$}},
  }{}\href{http://dx.doi.org/10.1103/PhysRevLett.114.132001}{Phys.\ Rev.\
  Lett.\  \textbf{114} (2014) 132001},
  \href{http://arxiv.org/abs/1411.2943}{{\normalfont\ttfamily
  arXiv:1411.2943}}\relax
\mciteBstWouldAddEndPuncttrue
\mciteSetBstMidEndSepPunct{\mcitedefaultmidpunct}
{\mcitedefaultendpunct}{\mcitedefaultseppunct}\relax
\EndOfBibitem
\bibitem{LHCb-PAPER-2014-056}
LHCb collaboration, R.~Aaij {\em et~al.},
  \ifthenelse{\boolean{articletitles}}{\emph{{Study of
  $\Peta$--$\Peta^{\prime}$ mixing from measurement of $\B^0_{(\squark)} \to
  \jpsi \Peta^{(\prime)}$ decay rates}},
  }{}\href{http://dx.doi.org/10.1007/JHEP01(2015)024}{JHEP \textbf{01} (2015)
  024}, \href{http://arxiv.org/abs/1411.0943}{{\normalfont\ttfamily
  arXiv:1411.0943}}\relax
\mciteBstWouldAddEndPuncttrue
\mciteSetBstMidEndSepPunct{\mcitedefaultmidpunct}
{\mcitedefaultendpunct}{\mcitedefaultseppunct}\relax
\EndOfBibitem
\bibitem{LHCb-PAPER-2015-024}
LHCb collaboration, R.~Aaij {\em et~al.},
  \ifthenelse{\boolean{articletitles}}{\emph{{Measurement of the branching
  fraction ratio $\mathcal{B}(\Bc\to\psitwos\pip)/\mathcal{B}(\Bc\to \jpsi
  \pip)$}}, }{}\href{http://dx.doi.org/10.1103/PhysRevD.92.072007}{Phys.\ Rev.\
   \textbf{D92} (2015) 057007},
  \href{http://arxiv.org/abs/1507.03516}{{\normalfont\ttfamily
  arXiv:1507.03516}}\relax
\mciteBstWouldAddEndPuncttrue
\mciteSetBstMidEndSepPunct{\mcitedefaultmidpunct}
{\mcitedefaultendpunct}{\mcitedefaultseppunct}\relax
\EndOfBibitem
\bibitem{AtlasLb}
\atlas collaboration, G.~Aad {\em et~al.},
  \ifthenelse{\boolean{articletitles}}{\emph{{Measurement of the branching
  ratio \mbox{$\Gamma(\Lb\to\psitwos\PLambda)/\Gamma(\Lb\to\jpsi\PLambda)$}
  with the \atlas detector}},
  }{}\href{http://dx.doi.org/10.1016/j.physletb.2015.10.009}{Phys.\ Lett.\
  \textbf{B751} (2015) 63},
  \href{http://arxiv.org/abs/1507.08202}{{\normalfont\ttfamily
  arXiv:1507.08202}}\relax
\mciteBstWouldAddEndPuncttrue
\mciteSetBstMidEndSepPunct{\mcitedefaultmidpunct}
{\mcitedefaultendpunct}{\mcitedefaultseppunct}\relax
\EndOfBibitem
\bibitem{Alves:2008zz}
LHCb collaboration, A.~A. Alves~Jr.\ {\em et~al.},
  \ifthenelse{\boolean{articletitles}}{\emph{{The \lhcb detector at the LHC}},
  }{}\href{http://dx.doi.org/10.1088/1748-0221/3/08/S08005}{JINST \textbf{3}
  (2008) S08005}\relax
\mciteBstWouldAddEndPuncttrue
\mciteSetBstMidEndSepPunct{\mcitedefaultmidpunct}
{\mcitedefaultendpunct}{\mcitedefaultseppunct}\relax
\EndOfBibitem
\bibitem{LHCb-DP-2014-002}
LHCb collaboration, R.~Aaij {\em et~al.},
  \ifthenelse{\boolean{articletitles}}{\emph{{LHCb detector performance}},
  }{}\href{http://dx.doi.org/10.1142/S0217751X15300227}{Int.\ J.\ Mod.\ Phys.\
  \textbf{A30} (2015) 1530022},
  \href{http://arxiv.org/abs/1412.6352}{{\normalfont\ttfamily
  arXiv:1412.6352}}\relax
\mciteBstWouldAddEndPuncttrue
\mciteSetBstMidEndSepPunct{\mcitedefaultmidpunct}
{\mcitedefaultendpunct}{\mcitedefaultseppunct}\relax
\EndOfBibitem
\bibitem{LHCb-DP-2014-001}
R.~Aaij {\em et~al.}, \ifthenelse{\boolean{articletitles}}{\emph{{Performance
  of the LHCb Vertex Locator}},
  }{}\href{http://dx.doi.org/10.1088/1748-0221/9/09/P09007}{JINST \textbf{9}
  (2014) P09007}, \href{http://arxiv.org/abs/1405.7808}{{\normalfont\ttfamily
  arXiv:1405.7808}}\relax
\mciteBstWouldAddEndPuncttrue
\mciteSetBstMidEndSepPunct{\mcitedefaultmidpunct}
{\mcitedefaultendpunct}{\mcitedefaultseppunct}\relax
\EndOfBibitem
\bibitem{LHCb-PAPER-2012-048}
LHCb collaboration, R.~Aaij {\em et~al.},
  \ifthenelse{\boolean{articletitles}}{\emph{{Measurements of the $\Lb$,
  $\Xi_{\bquark}^-$, and $\Omega_{\bquark}^-$ baryon masses}},
  }{}\href{http://dx.doi.org/10.1103/PhysRevLett.110.182001}{Phys.\ Rev.\
  Lett.\  \textbf{110} (2013) 182001},
  \href{http://arxiv.org/abs/1302.1072}{{\normalfont\ttfamily
  arXiv:1302.1072}}\relax
\mciteBstWouldAddEndPuncttrue
\mciteSetBstMidEndSepPunct{\mcitedefaultmidpunct}
{\mcitedefaultendpunct}{\mcitedefaultseppunct}\relax
\EndOfBibitem
\bibitem{LHCb-DP-2012-004}
R.~Aaij {\em et~al.}, \ifthenelse{\boolean{articletitles}}{\emph{{The \lhcb
  trigger and its performance in 2011}},
  }{}\href{http://dx.doi.org/10.1088/1748-0221/8/04/P04022}{JINST \textbf{8}
  (2013) P04022}, \href{http://arxiv.org/abs/1211.3055}{{\normalfont\ttfamily
  arXiv:1211.3055}}\relax
\mciteBstWouldAddEndPuncttrue
\mciteSetBstMidEndSepPunct{\mcitedefaultmidpunct}
{\mcitedefaultendpunct}{\mcitedefaultseppunct}\relax
\EndOfBibitem
\bibitem{Sjostrand:2006za}
T.~Sj\"{o}strand, S.~Mrenna, and P.~Skands,
  \ifthenelse{\boolean{articletitles}}{\emph{{PYTHIA 6.4 physics and manual}},
  }{}\href{http://dx.doi.org/10.1088/1126-6708/2006/05/026}{JHEP \textbf{05}
  (2006) 026}, \href{http://arxiv.org/abs/hep-ph/0603175}{{\normalfont\ttfamily
  arXiv:hep-ph/0603175}}\relax
\mciteBstWouldAddEndPuncttrue
\mciteSetBstMidEndSepPunct{\mcitedefaultmidpunct}
{\mcitedefaultendpunct}{\mcitedefaultseppunct}\relax
\EndOfBibitem
\bibitem{Sjostrand:2007gs}
T.~Sj\"{o}strand, S.~Mrenna, and P.~Skands,
  \ifthenelse{\boolean{articletitles}}{\emph{{A brief introduction to PYTHIA
  8.1}}, }{}\href{http://dx.doi.org/10.1016/j.cpc.2008.01.036}{Comput.\ Phys.\
  Commun.\  \textbf{178} (2008) 852},
  \href{http://arxiv.org/abs/0710.3820}{{\normalfont\ttfamily
  arXiv:0710.3820}}\relax
\mciteBstWouldAddEndPuncttrue
\mciteSetBstMidEndSepPunct{\mcitedefaultmidpunct}
{\mcitedefaultendpunct}{\mcitedefaultseppunct}\relax
\EndOfBibitem
\bibitem{LHCb-PROC-2010-056}
I.~Belyaev {\em et~al.}, \ifthenelse{\boolean{articletitles}}{\emph{{Handling
  of the generation of primary events in Gauss, the LHCb simulation
  framework}}, }{}\href{http://dx.doi.org/10.1088/1742-6596/331/3/032047}{{J.\
  Phys.\ Conf.\ Ser.\ } \textbf{331} (2011) 032047}\relax
\mciteBstWouldAddEndPuncttrue
\mciteSetBstMidEndSepPunct{\mcitedefaultmidpunct}
{\mcitedefaultendpunct}{\mcitedefaultseppunct}\relax
\EndOfBibitem
\bibitem{Lange:2001uf}
D.~J. Lange, \ifthenelse{\boolean{articletitles}}{\emph{{The EvtGen particle
  decay simulation package}},
  }{}\href{http://dx.doi.org/10.1016/S0168-9002(01)00089-4}{Nucl.\ Instrum.\
  Meth.\  \textbf{A462} (2001) 152}\relax
\mciteBstWouldAddEndPuncttrue
\mciteSetBstMidEndSepPunct{\mcitedefaultmidpunct}
{\mcitedefaultendpunct}{\mcitedefaultseppunct}\relax
\EndOfBibitem
\bibitem{Golonka:2005pn}
P.~Golonka and Z.~Was, \ifthenelse{\boolean{articletitles}}{\emph{{PHOTOS Monte
  Carlo: A precision tool for QED corrections in $Z$ and $W$ decays}},
  }{}\href{http://dx.doi.org/10.1140/epjc/s2005-02396-4}{Eur.\ Phys.\ J.\
  \textbf{C45} (2006) 97},
  \href{http://arxiv.org/abs/hep-ph/0506026}{{\normalfont\ttfamily
  arXiv:hep-ph/0506026}}\relax
\mciteBstWouldAddEndPuncttrue
\mciteSetBstMidEndSepPunct{\mcitedefaultmidpunct}
{\mcitedefaultendpunct}{\mcitedefaultseppunct}\relax
\EndOfBibitem
\bibitem{Allison:2006ve}
Geant4 collaboration, J.~Allison {\em et~al.},
  \ifthenelse{\boolean{articletitles}}{\emph{{Geant4 developments and
  applications}}, }{}\href{http://dx.doi.org/10.1109/TNS.2006.869826}{IEEE
  Trans.\ Nucl.\ Sci.\  \textbf{53} (2006) 270}\relax
\mciteBstWouldAddEndPuncttrue
\mciteSetBstMidEndSepPunct{\mcitedefaultmidpunct}
{\mcitedefaultendpunct}{\mcitedefaultseppunct}\relax
\EndOfBibitem
\bibitem{Agostinelli:2002hh}
Geant4 collaboration, S.~Agostinelli {\em et~al.},
  \ifthenelse{\boolean{articletitles}}{\emph{{Geant4: A simulation toolkit}},
  }{}\href{http://dx.doi.org/10.1016/S0168-9002(03)01368-8}{Nucl.\ Instrum.\
  Meth.\  \textbf{A506} (2003) 250}\relax
\mciteBstWouldAddEndPuncttrue
\mciteSetBstMidEndSepPunct{\mcitedefaultmidpunct}
{\mcitedefaultendpunct}{\mcitedefaultseppunct}\relax
\EndOfBibitem
\bibitem{LHCb-PROC-2011-006}
M.~Clemencic {\em et~al.}, \ifthenelse{\boolean{articletitles}}{\emph{{The
  \lhcb simulation application, Gauss: Design, evolution and experience}},
  }{}\href{http://dx.doi.org/10.1088/1742-6596/331/3/032023}{{J.\ Phys.\ Conf.\
  Ser.\ } \textbf{331} (2011) 032023}\relax
\mciteBstWouldAddEndPuncttrue
\mciteSetBstMidEndSepPunct{\mcitedefaultmidpunct}
{\mcitedefaultendpunct}{\mcitedefaultseppunct}\relax
\EndOfBibitem
\bibitem{LHCb-PAPER-2014-009}
LHCb collaboration, R.~Aaij {\em et~al.},
  \ifthenelse{\boolean{articletitles}}{\emph{{Evidence for the decay $\Bc \to
  \jpsi 3\pip2\pim$}}, }{}\href{http://dx.doi.org/10.1007/JHEP05(2014)148}{JHEP
  \textbf{05} (2014) 148},
  \href{http://arxiv.org/abs/1404.0287}{{\normalfont\ttfamily
  arXiv:1404.0287}}\relax
\mciteBstWouldAddEndPuncttrue
\mciteSetBstMidEndSepPunct{\mcitedefaultmidpunct}
{\mcitedefaultendpunct}{\mcitedefaultseppunct}\relax
\EndOfBibitem
\bibitem{LHCb-DP-2012-003}
M.~Adinolfi {\em et~al.},
  \ifthenelse{\boolean{articletitles}}{\emph{{Performance of the \lhcb RICH
  detector at the LHC}},
  }{}\href{http://dx.doi.org/10.1140/epjc/s10052-013-2431-9}{Eur.\ Phys.\ J.\
  \textbf{C73} (2013) 2431},
  \href{http://arxiv.org/abs/1211.6759}{{\normalfont\ttfamily
  arXiv:1211.6759}}\relax
\mciteBstWouldAddEndPuncttrue
\mciteSetBstMidEndSepPunct{\mcitedefaultmidpunct}
{\mcitedefaultendpunct}{\mcitedefaultseppunct}\relax
\EndOfBibitem
\bibitem{LHCb-DP-2013-001}
F.~Archilli {\em et~al.},
  \ifthenelse{\boolean{articletitles}}{\emph{{Performance of the muon
  identification at LHCb}},
  }{}\href{http://dx.doi.org/10.1088/1748-0221/8/10/P10020}{JINST \textbf{8}
  (2013) P10020}, \href{http://arxiv.org/abs/1306.0249}{{\normalfont\ttfamily
  arXiv:1306.0249}}\relax
\mciteBstWouldAddEndPuncttrue
\mciteSetBstMidEndSepPunct{\mcitedefaultmidpunct}
{\mcitedefaultendpunct}{\mcitedefaultseppunct}\relax
\EndOfBibitem
\bibitem{PDG2014}
Particle Data Group, K.~A. Olive {\em et~al.},
  \ifthenelse{\boolean{articletitles}}{\emph{{\href{http://pdg.lbl.gov/}{Review
  of particle physics}}},
  }{}\href{http://dx.doi.org/10.1088/1674-1137/38/9/090001}{Chin.\ Phys.\
  \textbf{C38} (2014) 090001}, {and 2015 update}\relax
\mciteBstWouldAddEndPuncttrue
\mciteSetBstMidEndSepPunct{\mcitedefaultmidpunct}
{\mcitedefaultendpunct}{\mcitedefaultseppunct}\relax
\EndOfBibitem
\bibitem{Hulsbergen:2005pu}
W.~D. Hulsbergen, \ifthenelse{\boolean{articletitles}}{\emph{{Decay chain
  fitting with a Kalman filter}},
  }{}\href{http://dx.doi.org/10.1016/j.nima.2005.06.078}{Nucl.\ Instrum.\
  Meth.\  \textbf{A552} (2005) 566},
  \href{http://arxiv.org/abs/physics/0503191}{{\normalfont\ttfamily
  arXiv:physics/0503191}}\relax
\mciteBstWouldAddEndPuncttrue
\mciteSetBstMidEndSepPunct{\mcitedefaultmidpunct}
{\mcitedefaultendpunct}{\mcitedefaultseppunct}\relax
\EndOfBibitem
\bibitem{Skwarnicki:1986xj}
T.~Skwarnicki, {\em {A study of the radiative cascade transitions between the
  $\Upsilon^{\prime}$ and $\Upsilon$ resonances}}, PhD thesis, Institute of
  Nuclear Physics, Krakow, 1986,
  {\href{http://inspirehep.net/record/230779/}{DESY-F31-86-02}}\relax
\mciteBstWouldAddEndPuncttrue
\mciteSetBstMidEndSepPunct{\mcitedefaultmidpunct}
{\mcitedefaultendpunct}{\mcitedefaultseppunct}\relax
\EndOfBibitem
\bibitem{LHCb-PAPER-2011-013}
LHCb collaboration, R.~Aaij {\em et~al.},
  \ifthenelse{\boolean{articletitles}}{\emph{{Observation of $\jpsi$-pair
  production in pp collisions at $\sqs=7\tev$}},
  }{}\href{http://dx.doi.org/10.1016/j.physletb.2011.12.015}{Phys.\ Lett.\
  \textbf{B707} (2012) 52},
  \href{http://arxiv.org/abs/1109.0963}{{\normalfont\ttfamily
  arXiv:1109.0963}}\relax
\mciteBstWouldAddEndPuncttrue
\mciteSetBstMidEndSepPunct{\mcitedefaultmidpunct}
{\mcitedefaultendpunct}{\mcitedefaultseppunct}\relax
\EndOfBibitem
\bibitem{Pivk:2004ty}
M.~Pivk and F.~R. Le~Diberder,
  \ifthenelse{\boolean{articletitles}}{\emph{{sPlot: A statistical tool to
  unfold data distributions}},
  }{}\href{http://dx.doi.org/10.1016/j.nima.2005.08.106}{Nucl.\ Instrum.\
  Meth.\  \textbf{A555} (2005) 356},
  \href{http://arxiv.org/abs/physics/0402083}{{\normalfont\ttfamily
  arXiv:physics/0402083}}\relax
\mciteBstWouldAddEndPuncttrue
\mciteSetBstMidEndSepPunct{\mcitedefaultmidpunct}
{\mcitedefaultendpunct}{\mcitedefaultseppunct}\relax
\EndOfBibitem
\bibitem{LHCb-DP-2013-002}
LHCb collaboration, R.~Aaij {\em et~al.},
  \ifthenelse{\boolean{articletitles}}{\emph{{Measurement of the track
  reconstruction efficiency at LHCb}},
  }{}\href{http://dx.doi.org/10.1088/1748-0221/10/02/P02007}{JINST \textbf{10}
  (2015) P02007}, \href{http://arxiv.org/abs/1408.1251}{{\normalfont\ttfamily
  arXiv:1408.1251}}\relax
\mciteBstWouldAddEndPuncttrue
\mciteSetBstMidEndSepPunct{\mcitedefaultmidpunct}
{\mcitedefaultendpunct}{\mcitedefaultseppunct}\relax
\EndOfBibitem
\bibitem{Lees:2011gw}
BaBar collaboration, J.~P. Lees {\em et~al.},
  \ifthenelse{\boolean{articletitles}}{\emph{{Branching fraction measurements
  of the~color\nobreakdash-suppressed decays \mbox{${\bar{\B}}^0 \to \D^{(*)0}
  \Ppi^0$}, \mbox{$\D^{(*)0} \Peta$}, \mbox{$\D^{(*)0} \Pomega$}, and
  \mbox{$\D^{(*)0} \Peta^\prime$} and measurement of the~polarization in
  the~decay \mbox{${\bar{\B}}^0\to \D^{*0} \Pomega$}}},
  }{}\href{http://dx.doi.org/10.1103/PhysRevD.84.112007}{Phys.\ Rev.\
  \textbf{D84} (2011) 112007}, Erratum
  \href{http://dx.doi.org/10.1103/PhysRevD.87.039901}{ibid.\   \textbf{D87}
  (2013) 039901}, \href{http://arxiv.org/abs/1107.5751}{{\normalfont\ttfamily
  arXiv:1107.5751}}\relax
\mciteBstWouldAddEndPuncttrue
\mciteSetBstMidEndSepPunct{\mcitedefaultmidpunct}
{\mcitedefaultendpunct}{\mcitedefaultseppunct}\relax
\EndOfBibitem
\bibitem{Apolonios}
D.~Martinez~Santos and F.~Dupertuis,
  \ifthenelse{\boolean{articletitles}}{\emph{{Mass distributions marginalized
  over per-event errors}},
  }{}\href{http://dx.doi.org/10.1016/j.nima.2014.06.081}{Nucl.\ Instrum.\
  Meth.\  \textbf{A764} (2014) 150},
  \href{http://arxiv.org/abs/1312.5000}{{\normalfont\ttfamily
  arXiv:1312.5000}}\relax
\mciteBstWouldAddEndPuncttrue
\mciteSetBstMidEndSepPunct{\mcitedefaultmidpunct}
{\mcitedefaultendpunct}{\mcitedefaultseppunct}\relax
\EndOfBibitem
\bibitem{BrPsi_E672}
E672 and E706 collaborations, A.~Gribushin {\em et~al.},
  \ifthenelse{\boolean{articletitles}}{\emph{{Production of $\jpsi$ and
  $\psitwos$ mesons in $\pim\PB\Pe$ collisions at $515\gevc$}},
  }{}\href{http://dx.doi.org/10.1103/PhysRevD.53.4723}{Phys.\ Rev.\
  \textbf{D53} (1996) 4723} FERMILAB-PUB-95-298-E\relax
\mciteBstWouldAddEndPuncttrue
\mciteSetBstMidEndSepPunct{\mcitedefaultmidpunct}
{\mcitedefaultendpunct}{\mcitedefaultseppunct}\relax
\EndOfBibitem
\bibitem{BrPsi_babar}
\babar collaboration, B.~Aubert {\em et~al.},
  \ifthenelse{\boolean{articletitles}}{\emph{{Measurement of the branching
  fractions for $\psitwos\to\Pe^+\Pe^-$ and $\psimm$}},
  }{}\href{http://dx.doi.org/10.1103/PhysRevD.65.031101}{Phys.\ Rev.\
  \textbf{D65} (2002) 031101},
  \href{http://arxiv.org/abs/hep-ex/0109004}{{\normalfont\ttfamily
  arXiv:hep-ex/0109004}}\relax
\mciteBstWouldAddEndPuncttrue
\mciteSetBstMidEndSepPunct{\mcitedefaultmidpunct}
{\mcitedefaultendpunct}{\mcitedefaultseppunct}\relax
\EndOfBibitem
\bibitem{LHCb-PAPER-2013-011}
LHCb collaboration, R.~Aaij {\em et~al.},
  \ifthenelse{\boolean{articletitles}}{\emph{{Precision measurement of $\D$
  meson mass differences}},
  }{}\href{http://dx.doi.org/10.1007/JHEP06(2013)065}{JHEP \textbf{06} (2013)
  065}, \href{http://arxiv.org/abs/1304.6865}{{\normalfont\ttfamily
  arXiv:1304.6865}}\relax
\mciteBstWouldAddEndPuncttrue
\mciteSetBstMidEndSepPunct{\mcitedefaultmidpunct}
{\mcitedefaultendpunct}{\mcitedefaultseppunct}\relax
\EndOfBibitem
\bibitem{LHCb-PAPER-2011-035}
LHCb collaboration, R.~Aaij {\em et~al.},
  \ifthenelse{\boolean{articletitles}}{\emph{{Measurement of $\bquark$-hadron
  masses}}, }{}\href{http://dx.doi.org/10.1016/j.physletb.2012.01.058}{Phys.\
  Lett.\  \textbf{B708} (2012) 241},
  \href{http://arxiv.org/abs/1112.4896}{{\normalfont\ttfamily
  arXiv:1112.4896}}\relax
\mciteBstWouldAddEndPuncttrue
\mciteSetBstMidEndSepPunct{\mcitedefaultmidpunct}
{\mcitedefaultendpunct}{\mcitedefaultseppunct}\relax
\EndOfBibitem
\bibitem{LHCb-PAPER-2010-001}
LHCb collaboration, R.~Aaij {\em et~al.},
  \ifthenelse{\boolean{articletitles}}{\emph{{Prompt
  $\mathrm{K}^0_{\mathrm{S}}$ production in $\proton\proton$ collisions at
  $\sqs=0.9\tev$}},
  }{}\href{http://dx.doi.org/10.1016/j.physletb.2010.08.055}{Phys.\ Lett.\
  \textbf{B693} (2010) 69},
  \href{http://arxiv.org/abs/1008.3105}{{\normalfont\ttfamily
  arXiv:1008.3105}}\relax
\mciteBstWouldAddEndPuncttrue
\mciteSetBstMidEndSepPunct{\mcitedefaultmidpunct}
{\mcitedefaultendpunct}{\mcitedefaultseppunct}\relax
\EndOfBibitem
\bibitem{LbMassAtlas}
\atlas collaboration, G.~Aad {\em et~al.},
  \ifthenelse{\boolean{articletitles}}{\emph{{Measurement of the \Lb lifetime
  and mass in the \atlas experiment}},
  }{}\href{http://dx.doi.org/10.1103/PhysRevD.87.032002}{Phys.\ Rev.\
  \textbf{D87} (2013) 032002},
  \href{http://arxiv.org/abs/1207.2284}{{\normalfont\ttfamily
  arXiv:1207.2284}}\relax
\mciteBstWouldAddEndPuncttrue
\mciteSetBstMidEndSepPunct{\mcitedefaultmidpunct}
{\mcitedefaultendpunct}{\mcitedefaultseppunct}\relax
\EndOfBibitem
\end{mcitethebibliography}

\newpage
%%%%%%%%%%%%%%%%%%%%%%%%%%%%%%%%%%%%%%%%%%
\centerline{\large\bf LHCb collaboration}
\begin{flushleft}
\small
R.~Aaij$^{39}$, 
C.~Abell\'{a}n~Beteta$^{41}$, 
B.~Adeva$^{38}$, 
M.~Adinolfi$^{47}$, 
A.~Affolder$^{53}$, 
Z.~Ajaltouni$^{5}$, 
S.~Akar$^{6}$, 
J.~Albrecht$^{10}$, 
F.~Alessio$^{39}$, 
M.~Alexander$^{52}$
% ...

{\footnotesize \it
% ...
$ ^{5}$Clermont Universit\'{e}, Universit\'{e} Blaise Pascal, CNRS/IN2P3, LPC, Clermont-Ferrand, France\\
$ ^{6}$CPPM, Aix-Marseille Universit\'{e}, CNRS/IN2P3, Marseille, France\\
% ...
$ ^{10}$Fakult\"{a}t Physik, Technische Universit\"{a}t Dortmund, Dortmund, Germany\\
% ...
$ ^{38}$Universidad de Santiago de Compostela, Santiago de Compostela, Spain\\
$ ^{39}$European Organization for Nuclear Research (CERN), Geneva, Switzerland\\
% ...
$ ^{41}$Physik-Institut, Universit\"{a}t Z\"{u}rich, Z\"{u}rich, Switzerland\\
% ...
$ ^{47}$H.H. Wills Physics Laboratory, University of Bristol, Bristol, United Kingdom\\
% ...
$ ^{52}$School of Physics and Astronomy, University of Glasgow, Glasgow, United Kingdom\\
$ ^{53}$Oliver Lodge Laboratory, University of Liverpool, Liverpool, United Kingdom\\
% ...
}
\end{flushleft}
%%%%%%%%%%%%%%%%%%%%%%%%%%%%%%%%%%%%%%%%%%

\end{document}

%%%%%%%%%%%%%%%%%%%%%%%%%
%%%%% Title     %%%%%%%%%
%%%%%%%%%%%%%%%%%%%%%%%%%
\renewcommand{\thefootnote}{\fnsymbol{footnote}}
\setcounter{footnote}{1}

\renewcommand{\thefootnote}{\arabic{footnote}}
\setcounter{footnote}{0}

%%%%%%%%%%%%%%%%%%%%%%%%%%%%%%%%
%%%%%  Table of Content   %%%%%%
%%%%%%%%%%%%%%%%%%%%%%%%%%%%%%%%
%%%% Uncomment next 2 lines if desired
%\tableofcontents
%\cleardoublepage

%%%%%%%%%%%%%%%%%%%%%%%%%
%%%%% Main text %%%%%%%%%
%%%%%%%%%%%%%%%%%%%%%%%%%

\pagestyle{plain} % restore page numbers for the main text
\setcounter{page}{1}
\pagenumbering{arabic}

%% Uncomment during review phase. 
%% Comment before a final submission.
\linenumbers

% You can include short sections directly in the main tex file.
% However, for larger papers it is desirable to split the text into
% several semiautonomous files, which can be revised independently.
% This is especially useful when developing a document in
% collaboration with several people, since then different parts can be
% edited independently.  This type of file organization is shown here.

\section*{Supplementary material for LHCb-PAPER-2015-060}

%%%%%%%%%%%%%%%%%%%%%%%%%%%%%%%%%%%%%%%%%%%%%%%%%%%%%%%%%%%%%%%%%%%%%

\subsection*{Ratios of branching fractions of $\bquark$ hadron decays to $\psitwos$ and $\jpsi$}

\begin{figure}[h]
  \setlength{\unitlength}{1mm}
  \centering
   \begin{picture}(100,140)
    \put(0,0){\includegraphics*[width=100mm]{Fig1_supp.pdf}} % BrB2psiComp2.pdf
   \end{picture}
   \caption { \small Ratios of branching fractions of $\bquark$ hadron decays to final states
     containing $\psitwos$ and $\jpsi$.
     Red circles correspond to measurements performed by \lhcb collaboration~\cite{LHCb-PAPER-2015-060,LHCb-PAPER-2015-024,LHCb-PAPER-2012-054,LHCb-PAPER-2012-053,LHCb-PAPER-2014-056,LHCb-PAPER-2012-010,LHCb-PAPER-2011-024}
     while black squares represent measurements of the other experiments~\cite{BrB2psiX_CDF_1998,BrB2psiX_CDF_2006,BrB2psiX_D0_2009}.
     Black outer error bars represent the total uncertainties of the measurements,
     while the blue inner error bars represent statistical uncertainties only. }
   \label{fig:BrPsiAverage1}
\end{figure}

%%%%%%%%%%%%%%%%%%%%%%%%%%%%%%%%%%%%%%%%%%%%%%%%%%%%%%%%%%%%%%%%%%%%%

\newpage
\subsection*{\psitwos branching fractions}

\begin{figure}[h]
  \setlength{\unitlength}{1mm}
  \centering
   \begin{picture}(100,110)
    \put(0,0){\includegraphics*[width=100mm]{Fig2_supp.pdf}}
  \end{picture}
  \caption { \small Comparison of a ratio of the $\psimm$ and $\psijpp$ branching fractions measured in this paper (red star) with previous results of 
E672/E706~\cite{BrPsi_E672} %%~\footnote{Phys. Rev. {\bf D53} (1996) 4723}  %%~\cite{BrPsi_E672} 
and \babar~\cite{BrPsi_babar} %%~\footnote{Phys. Rev. {\bf D65} (2002) 031101} %%~\cite{BrPsi_babar} 
collaborations (black circles) and average and fit values derived by Particle Data Group~\cite{PDG2014} 
(green squares). The third green square represents the fit value of the Particle Data Group for the ratio of $\psitwos\to\Pe^+\Pe^-$ and $\psijpp$ branching 
fractions. Black and green outer error bars represent the total uncertainties of the measurements, while the blue inner error bars represent statistical 
uncertainties only. }
  \label{fig:BrPsiAverage2}
\end{figure}

%%%%%%%%%%%%%%%%%%%%%%%%%%%%%%%%%%%%%%%%%%%%%%%%%%%%%%%%%%%%%%%%%%%%%

\newpage
\subsection*{\lhcb average of the \Lb mass}

\begin{figure}[h]
  \setlength{\unitlength}{1mm}
  \centering
   \begin{picture}(100,140)
    \put(0,0){\includegraphics*[width=100mm]{Fig3_supp.pdf}} %MassAverage.pdf
  \end{picture}
  \caption { \small Comparison of the measurements of $\Lb$ mass of this analysis\,(black circles)
           with the previous results using ${\Lb\to\jpsi\Lambda}$ decay
           in 2010~\cite{LHCb-PAPER-2011-035} and in 2011~\cite{LHCb-PAPER-2012-048} data\,(black squares)
           and \Lb mass derived from the~known $\Bd$ mass and the~\mbox{$\Lb$--$\Bd$} 
           mass difference measured with ${\Lb\to\Lc\Dsm}$ decay in 2011 data~\cite{LHCb-PAPER-2014-002}\,(black diamond). 
           Black outer error bars represent the total uncertainties of the measurements,
           while the~blue inner error bars represent statistical uncertainties only.
           The~green points represent the averages of previous \lhcb measurements and measurements of this analysis,
           while the~red point is the~average of all the
           measurements excluding the one using the ${\Lb\to\Lc\Dsm}$ decay,
           which is in fact a measurement of $\Lb$-$\Bd$ mass difference.}
  \label{fig:MassAverage}
\end{figure}

%%%%%%%%%%%%%%%%%%%%%%%%%%%%%%%%%%%%%%%%%%%%%%%%%%%%%%%%%%%%%%%%%%%%%

\newpage
\subsection*{Measurements of the \Lb mass}

\begin{figure}[h]
  \setlength{\unitlength}{1mm}
  \centering
   \begin{picture}(160,140)
    \put(0,0){\includegraphics*[width=160mm]{lamdb.pdf}}
  \end{picture}
  \caption { \small Comparison of the measurements of $\Lb$ mass in
           the~\atlas~\cite{LbMassAtlas}, \cdf~\cite{LbMassCDF} and \lhcb~experiments. }
  \label{fig:MassWorldAverage}
\end{figure}

%%%%%%%%%%%%%%%%%%%%%%%%%%%%%%%%%%%%%%%%%%%%%%%%%%%%%%%%%%%%%%%%%%%%%

\newpage
\addcontentsline{toc}{section}{References}
\setboolean{inbibliography}{true}
\bibliographystyle{LHCb}
\bibliography{main,local,LHCb-PAPER,LHCb-CONF,LHCb-DP,LHCb-TDR}